\documentclass[nonacm,manuscript]{acmart}
\usepackage[english]{babel}
\usepackage{pifont}
\usepackage{xspace}
\usepackage{multirow,makecell}
\usepackage{tabularx}
\AtBeginDocument{%
  }
    
\newcounter{pitfallc}
\newcommand{\pitfall}[2]{%
    \refstepcounter{pitfallc}%
  \label{#2}%
  \paragraph{\textbf{Pitfall~\thepitfallc: #1}}%
}
\newcommand{\pref}[1]{Pitfall~\ref{#1}}%

\newcounter{solutionc}
\newcommand{\solution}[2]{%
    \refstepcounter{solutionc}%
  \label{#2}%
  \paragraph{\textbf{Solution~\thesolutionc: #1}}%
}
\newcommand{\sref}[1]{Solution~\ref{#1}}%

\setcopyright{none}
\copyrightyear{2025}
\acmYear{2025}

\usepackage[nameinlink]{cleveref}

\usepackage{svg}

\begin{document}
\title{Point of Interest Recommendation: Pitfalls and Viable Solutions}

\author{Alejandro Bellogín}
\authornote{All authors contributed equally to this research and are listed in alphabetical order.
}
\email{alejandro.bellogin@uam.es}
\orcid{0000-0001-6368-2510}
\affiliation{%
  \institution{Universidad Autónoma de Madrid}
  \city{Madrid}
  \country{Spain}}
  
\author{Linus W. Dietz}
\email{linus.dietz@kcl.ac.uk}
\orcid{0000-0001-6747-3898}
\affiliation{%
  \institution{King’s College London}
  \city{London}
  \country{United Kingdom}
}

\author{Francesco Ricci}
\email{fricci@unibz.it}
\orcid{0000-0001-5931-5566}
\affiliation{%
  \institution{Free University of Bozen-Bolzano}
  \city{Bolzano}
  \country{Italy}}

\author{Pablo Sánchez}
\email{psperez@icai.comillas.edu}
\orcid{0000-0003-1792-1706}
\affiliation{%
  \institution{Instituto de Investigación Tecnológica (IIT), Universidad Pontificia Comillas}
  \city{Madrid}
  \country{Spain}
}

\renewcommand{\shortauthors}{Bellogín, Dietz, Ricci, \& Sánchez}

\newcommand{\LBSNs}{Location-Based Social Networks\xspace}
\newcommand{\LBSNsAbbr}{LBSNs\xspace}

\begin{abstract}
Point of interest (POI) recommendation can play a pivotal role in enriching tourists' experiences by suggesting context-dependent and preference-matching locations and activities, such as restaurants, landmarks, itineraries, and cultural attractions.
Unlike some more common recommendation domains (e.g., music and video), POI recommendation is inherently high-stakes: users invest significant time, money, and effort to search, choose, and consume these suggested POIs. Despite the numerous research works in the area, several fundamental issues remain unresolved, hindering the real-world applicability of the proposed approaches. In this paper, we discuss the current status of the POI recommendation problem and the main challenges we have identified.

The first contribution of this paper is a critical assessment of the current state of POI recommendation research and the identification of key shortcomings across three main dimensions: datasets, algorithms, and evaluation methodologies. We highlight persistent issues such as the lack of standardized benchmark datasets, flawed assumptions in the problem definition and model design, and inadequate treatment of biases in the user behavior and system performance. 
The second contribution is a structured research agenda that, starting from the identified issues, introduces important directions for future work related to multistakeholder design, context awareness, data collection, trustworthiness, novel interactions, and real-world evaluation.
\end{abstract}


\maketitle

\section{Introduction}
\label{sec:intro}

Tourism offers a rich and multifaceted environment for the development of recommender systems (RSs). Assisting tourists in processing information and making decisions is inherently tied to the economic, social, and environmental dimensions of the destinations they visit. In this reflection paper, we focus on point of interest (POI) recommender technologies, and we critically assess the state of the art in this field, with the final target of overcoming the identified limitations that hinder the real-world adoption of recommender systems in this domain.
Our goal is to outline how POI recommender systems can be developed so they help travelers in their personalized decision-making process, but also considering the long-term sustainability and development goals of the places they wish to experience%
~\cite{ROBINARAMIREZ202412639,Banerjee2020}. As recommender systems become more common in daily life, POI recommendation is no exception. POI features are built into mobile apps for navigation, travel planning, and booking. Based on past behavior of users, these systems help them find landmarks, restaurants, museums, and local hidden gems, making it easier to explore unfamiliar cities.

However, tourism is different from traditional recommendation domains such as books or movies~\cite{Burke2011}. The stakes are higher as users invest time, money, and emotional effort to find these experiences, and the success of a recommendation depends on a combination of contextual factors: the atmosphere of a venue, the weather~\cite{Trattner2018Investigating,BraunhoferE0S14}, the composition of the travel group~\cite{Song2025Accurate,DBLP:journals/umuai/AlvesMSCNM23}, and more~\cite{jannach2020interactive,Herzog2019,leal2018context}. These complexities, combined with the global scale of tourism, make it hard to design effective recommendation algorithms, ensure smooth user-system interaction, and evaluate systems through experiments that are both rigorous and broadly applicable. As a result, much of the research in this area often overlooks real user needs, and the reported results are difficult to reproduce, which limits adoption in production systems.

In this reflection paper, we aim to disentangle persistent shortcomings in the research on POI recommender systems. 
In particular, our goal is to improve best practices in the academic community by critically examining common problems across three interconnected dimensions: datasets, algorithms, and evaluation methodologies.
We identify 20 pitfalls that, based on our review of the literature and discussions on the state of the art, are among the most prevalent and critical. The list is not comprehensive and is not intended as a criticism of individual studies. Rather, it should be read as a call to action to strengthen the impact and rigor of research in this field. 

When it comes to datasets, we find that much of the current research, especially offline evaluation of recommendation models, relies on outdated, incomplete, or biased data sourced from Location-Based Social Networks (LBSNs). User studies can offer more detailed insights; however, they remain underutilized and are often constrained by scalability challenges. This imbalance limits our understanding of the behavior of real-world users.

POI recommendation algorithms often prioritize accuracy metrics, which, while important, can overshadow other goals such as contextual relevance, diversity, and fairness.  
This narrow focus tends to favor popular places that are easier to predict, reducing personalization and limiting discovery. 
Recommending obvious, well-known places adds little value; travelers seeking popular spots likely do not need a recommender. What makes POI recommender systems meaningful is their potential to foster exploration, i.e., to help users discover venues they are likely to enjoy but might not have found on their own~\cite{Paromita23}.

Finally, we take a critical look at how POI recommender systems are evaluated. We argue that too many studies rely on narrow metrics and artificial testing environments, failing to reflect the rich complexity of tourist behavior in real-world settings. The lack of standardized data pre-processing strategies further leads to experimental outcomes that are frequently detached from the dynamic and context-sensitive nature of travel. Furthermore, we critically address incremental improvements and challenges in the reproducibility of algorithmic performance~\cite{Dacrema2021}.

Starting from the identified issues, as a second contribution, we propose a research agenda that states important directions for future work in this area that are related to multistakeholder design, context awareness, data collection, trustworthiness, novel interactions, and real-world evaluation. This research agenda includes the adoption of richer and more inclusive benchmark datasets, together with models that adapt to shifting user intentions, social group dynamics, and real-time constraints. Furthermore, the POI recommendation field needs to adopt more informative evaluation procedures and frameworks that are transparent about data pre-processing and emphasize meaningful, multistakeholder outcomes over accuracy-driven benchmarks.

The remainder of the paper is structured as follows. 
\Cref{sec:problem} introduces the application problem, stating the specific characteristics and challenges of POI recommendation in tourism contexts. 
\Cref{sec:soa} briefly reviews the state of the art, summarizing the main approaches and research topics in the field. 
\Cref{s:data_issues} discusses the limitations and open issues related to the datasets commonly used in POI recommendation works. 
\Cref{sec:alg} focuses on algorithmic limitations, highlighting aspects such as evaluation biases, lack of contextual modeling, and ignoring multistakeholders, while
\Cref{sec:eval} addresses widespread shortcomings in evaluation practices.
\Cref{sec:agenda} presents our research agenda to address the previously mentioned issues. 
Finally, \Cref{sec:conc} summarizes the lessons learned and concludes the paper.

\begin{figure}[tb]
    \centering
    \includegraphics[width=0.9\linewidth]{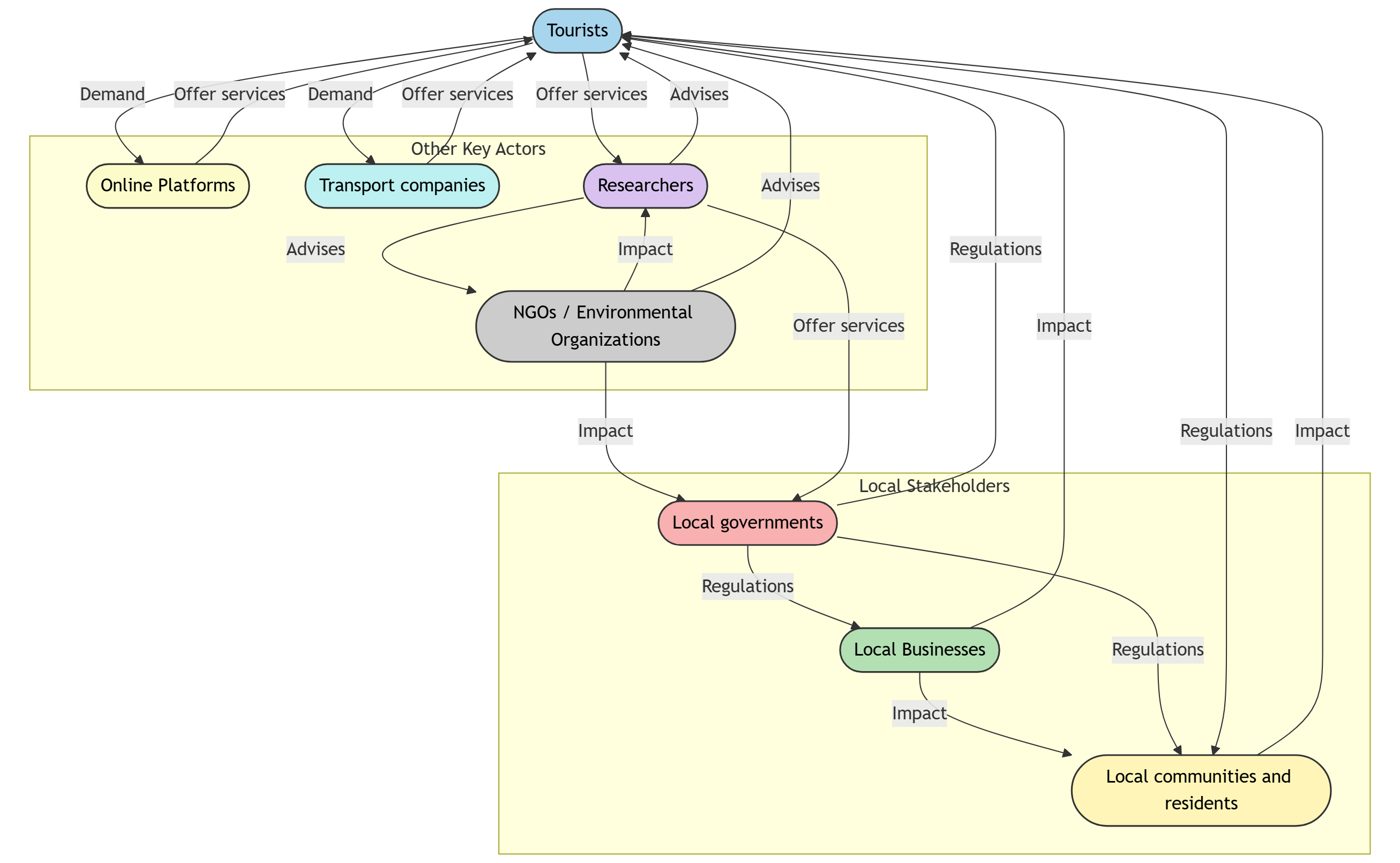}
    \caption{Relationships between various entities concerning regional sustainability, tourism, and the environment, based on \cite{Werthner01011999,pgr,10.1007/978-3-031-23844-4_26}.}
    \label{fig:stakeholders}
    \Description{TODO}
\end{figure}

\section{Application Problem}
\label{sec:problem}
Point of interest recommendations can play a crucial role in helping travelers discover relevant venues and better organize their visits throughout the different stages of their trip, assisting them in the whole decision-making process. Properly defining the POI recommendation problem requires differentiating between tourist information processing and decision-making activities that are conducted before travel (\textit{pre-trip planning}), and during travel to the selected destination (\textit{in-trip planning}) \cite{DBLP:journals/expert/StaabWRZGFPK02}. In both scenarios, it is important to remember that tourists have diverse needs and limits. They may prefer certain types of venues, face time or budget constraints, seek novel and varied suggestions, or need to meet the preferences of others in their group. These needs and interests affect the choices they make and are often influenced by promotion activities performed by a wide range of stakeholders in the tourism market, such as destination management organizations (DMOs), local governments, booking platforms, tour operators, airline companies, travel agencies, or even social network influencers~\cite{Werthner01011999,pgr,10.1007/978-3-031-23844-4_26}, as depicted in \Cref{fig:stakeholders}.  Each of these stakeholders has its own economic, social, and political objectives, which could be complementary but also contrary to the interests of the tourists.
Hence, the tourism recommendation scenario must be understood as a \textit{multistakeholder scenario}~\cite{Boom16022021,Merinov024,AbdollahpouriB22} and an effective recommender system must find the right balance strategy when performing POI recommendations. 

Moreover, while tourists search for personalized and satisfying experiences, %
today, DMOs are also influenced by local communities to ensure a more sustainable development. That is, they aim to increase economic growth while preserving social and environmental resources. For example, %
DMOs are looking for strategies to promote lesser-known areas and reduce the overcrowding of popular venues \cite{pechlaner2019overtourism,smeral2019overcrowding,pechlaner2000cultural,khalil2021monitoring,perry2023role}. 
This is also motivated by local businesses that look for fair exposure to attract visitors and compete with the few already popular POIs. %
As we have shown, the implementation of POI recommender systems is especially complex and delicate, due to all the conflicting objectives of the different stakeholders involved in the process. Consequently, while designing RSs and evaluating the quality of the generated recommendations, it is not enough to simply match user preferences with recommendations. It is necessary to study other %
indicators, including travelers' satisfaction, business development, spatial dispersion of tourists, local community well-being, or sustainability~\cite{DBLP:conf/um/PiliponyteM023}.

In addition to these aspects that define the POI recommendation application problem, it is necessary to take into account that tourism is often experienced as a group activity. Families, friends, couples, or ad hoc organized groups plan their visits and experience together the chosen POIs. This dimension increases the complexity of the recommendation task, as within a group, its members can exhibit divergent preferences and constraints~\cite{JamesonWF22}. 
Developing tourism recommender systems that consider both the potentially conflicting objectives and the dynamics of group decision-making is necessary to generate fair and inclusive recommendations to satisfy all group members.

While all of these aspects should be considered in both \textit{pre-trip planning} and \textit{in-trip planning} scenarios, each of them also requires distinct functionalities and user interfaces, as they follow different goals and contextual constraints.
On the one hand, 
\textit{pre-trip planning} is more related to high-level selection of which cities or regions to visit and what activities to perform there. It is performed over a longer time frame, sometimes weeks and months, and the user and the system exchange and process more information.
In contrast, \textit{in-trip planning} is strongly influenced by the real-time context, including POI availability, distance to the POI, time and weather constraints, group dynamics, and evolving preferences. %
In fact, at the core of the first scenario lies the notion of \textit{destination}, which is not only a geographical place. It is a subjective cultural concept defined by the cultural background of the traveler, the prior knowledge of the destination, and the distance of the traveler to the destination~\cite{DBLP:conf/rectour/GoldenbergMHKLM21,Dietz2022a}.
A destination is also shaped by tourist behavior. Places or activities not initially seen as appealing may become attractive due to factors such as content shared on social media, tourism websites, and other communication platforms. In addition, targeted media campaigns and the spread of information can influence visitors' interest and attention. The reasons for visiting a given POI vary by tourist and are shaped by factors such as cultural background, prior knowledge, and specific travel goals—whether to study, relax, satisfy curiosity, or something else~\cite{doi:10.1177/001088047401400409,gnoth1997tourism,pearce2005developing}. 

Therefore, RSs designed to support \emph{pre-trip planning} must go beyond traditional information filtering. Rather than simply selecting items, they should act as intermediaries, offering personalized content that matches the different cultural and emotional values of the destination with the traveler's interest~\cite{DBLP:journals/expert/StaabWRZGFPK02}. Hence, effective POI RSs should not only recommend relevant POIs but also show that the recommendations satisfy the user's underlying motivations and preferences through justified descriptions or explanations~\cite{Tintarev2022Beyond}. It is important to understand how to segment the venues of a larger territory, how to represent them to better match the wide variety of tourists' needs, and how to properly describe how each POI could match the general user preferences~\cite{HERNANDEZ201835,paulino2021identifying}. 
For these reasons, the role of RSs in POI recommendation is expanding. They are becoming moderators of information exchange between suppliers and consumers to identify underused areas, suggest promotional opportunities for emerging places, and support marketing strategies based on traveler demand and behavior, especially when the destination is new to the user.

Focusing now on the second scenario (\textit{in-trip planning}), which includes the real-time decision-making processes that tourists perform when they \textit{are} visiting a selected destination, it is clear that the functionality of the RSs must be distinct from that suitable for the pre-trip planning. 
The core factors to consider here are the contextual and dynamic information available at the time of the visit, as they determine the relevance and feasibility of visiting each POI, based on the user's current situation, environmental conditions, and POI-specific constraints. This contextual information falls into three main categories.
First, \emph{POI-related context} includes factors such as distance, estimated travel and visit times, opening hours, and ongoing events.
Second, \emph{user-specific context} refers to the traveler’s current interests, needs, and constraints, such as fatigue, hunger, mood, or group preferences.
Third, \emph{external context} includes factors not directly tied to the tourist but still influential, such as weather, air pollution, temporary disruptions, urban noise, or crowd density.
The combination of these three types of contextual factors can significantly influence tourists' decision-making process, potentially leading to changes in their original planned itinerary. For example, a group may plan to visit two museums in a row, but after the first visit, they decide to rest at a nearby bar before continuing.

\section{State of the Art} %
\label{sec:soa}

In this section, we briefly survey some important results from the state of the art of POI recommendation. We structure the description in three key dimensions: data, algorithms, and system evaluation. The goal is to establish the basis for identifying research gaps and solutions that are the contributions of this work. This overview is not comprehensive, and we refer readers to existing surveys for a more detailed analysis~\cite{DBLP:journals/csur/SanchezB22, DBLP:journals/jitt/WangHJ25}.

\subsection{Data Used in POI Recommendation}
\label{ss:Data}

Datasets based on LBSNs such as Foursquare, Gowalla, or Yelp, have been widely used in the POI recommendation literature as they provide large-scale check-in data from users with diverse profiles, including different types of tourists and locals~\cite{DBLP:journals/csur/SanchezB22}. In some cases, these LBSNs also provide information regarding the POIs such as the type of establishment and reviews, as well as user data, including gender and followers. However, despite their usefulness and usage, these datasets present some limitations that must also be considered and will be discussed in \Cref{s:data_issues}.
In addition to LBSNs, some works use photo-based platforms such as Flickr, where users store, organize, and publicly share POI images, together with rich metadata. TripBuilder~\cite{DBLP:conf/cikm/BrilhanteMNPR13} %
is an example of a dataset based on Flickr used in tourism RSs: it includes reconstructed POI visit trajectories in three major Italian touristic cities.
These data logs were created in a fully automated process by using geo-tagged photos from Flickr, and subsequently identifying the featured POIs and gathering additional information by using Wikipedia. This Flickr data was also used in~\cite{Jarv17}, where the authors dealt with errors in both timestamps and geographical coordinates of the originally extracted trajectories. This issue was also pointed out by~\cite{Dietz2020}. However, Flickr data provides good indicators of the spatial regions preferred by tourists and POI popularity, by assuming that photo sharing reflects user interest. %
Among the publicly available Flickr datasets, the Yahoo Flickr Creative Commons 100 Million Dataset (YFCC100M) \cite{DBLP:journals/cacm/ThomeeSFENPBL16} stands out for its scale and usage. It has served as the basis for multiple derived datasets in tourism and POI recommendation~\cite{DBLP:conf/aips/LimCLK16, DBLP:journals/kais/LimCLK18}.

Other recent data sources are emerging outside the domain of LBSNs. For example, the YJMob100K dataset~\cite{DBLP:data/10/YabeTSSSMP24} %
contains large-scale anonymized human mobility trajectories collected from 100,000 mobile phone users over 90 days, covering both regular and emergency scenarios. This dataset, originally intended for the prediction of urban mobility, may offer promising opportunities to evaluate POI and trajectory recommendation models under realistic and changing conditions. 
Tourist-oriented data can be much less noisy when interactions with POIs are explicitly recorded, for example, through visit actions with visitor cards. While such data is rare, a notable case is when tourists log their activities using a visitor card or city pass, as was done in Verona~\cite{DBLP:journals/tetc/MiglioriniCB21}.

Finally, when real data is scarce or incomplete, synthetic datasets have been proposed to fill this gap. For example, \citeauthor{MerinovM023} propose a mechanism that generates individual-level user profiles from aggregated population-level tourism data, and simulates POI visits using a discrete choice model with divergence minimization~\cite{MerinovM023}. 
A recent trend in the community for creating synthetic datasets consists of using large language models (LLMs) to emulate users. In this case, instead of generating full trajectories or proposing a new recommendation algorithm, \citeauthor{banerjee2025synthtrips} focus on generating realistic and diverse travel queries used by a recommender based on LLMs~\cite{banerjee2025synthtrips}.

\subsection{Algorithms and Recommendation Models}
\label{s:algorithms}

Researchers have proposed a wide variety of algorithmic approaches in the development of tourism and POI recommender systems, ranging from traditional 
nearest neighbors 
and matrix factorization techniques to more recent strategies based on neural networks. Classical models--especially those based on similarities--are often used because of their interpretability and simplicity. These approaches, together with matrix factorization, try to predict user preferences based on previous check-ins or ratings by identifying similar patterns or latent interactions in the user-POI interaction or rating matrix. However, the use of these techniques has been somewhat displaced due to advances in neural network-based architectures, although there are recent hybrid proposals that use 
nearest neighbors
or matrix factorization algorithms as part of their model~\cite{DBLP:journals/csur/SanchezB22}.

In recent years, the rapid progress of LLMs has gained attention in both tourism and POI recommendation. For example, \citeauthor{DBLP:conf/sigir/LiR0A0S24} address the next-POI recommendation problem by proposing a new method that uses pretrained language models to better exploit the rich contextual data from LBSNs~\cite{DBLP:conf/sigir/LiR0A0S24}. Unlike previous approaches that rely on numerical representations, the proposed framework maintains the original structure of the data and relies on contextual information.
\citeauthor{DBLP:conf/recsys/WangX0GCX24} present SeCor~\cite{DBLP:conf/recsys/WangX0GCX24}, a new approach to next-POI recommendation that combines collaborative and semantic information using a multi-modal strategy. It addresses limitations of previous models in capturing complex spatio-temporal patterns, while also mitigating LLM-induced hallucinations. By combining user–POI interaction embeddings with rich semantic representations, their model generates more robust and expressive hybrid encodings that improve recommendation performance.

Reinforcement learning has also been adopted as a promising solution to the challenges of POI recommendation, such as the cold-start problem. \citeauthor{Massimo023} introduce a reinforcement learning tourism RS called \emph{``QEXP''} that uses a model based on POI visits behavior extracted from POI visits logs~\cite{Massimo023}.  A related approach introduces a hierarchical reinforcement learning framework named \emph{``HRL-PRP''} designed to improve POI recommendation by better capturing complex user behavior~\cite{DBLP:conf/ijcai/Xiao00XWY24}. In contrast to traditional models, their proposal revises user profiles through a two-level decision process: the high-level component decides whether the user's profile should be updated, while the low-level component selects which noisy POIs from the user's history should be removed.

Cross-domain and transfer learning approaches have also been used to mitigate sparsity and improve recommendations. 
For instance, \citeauthor{DBLP:conf/www/ZhengZXY10} exploited information from 
GPS %
logs, POI databases, and the Web to obtain location-activity, location-feature, and activity-activity correlations to improve recommendation performance~\cite{DBLP:conf/www/ZhengZXY10}, whereas \citeauthor{DBLP:journals/ipm/SanchezB21} demonstrate that augmenting data with nearby cities' check-ins improves accuracy in cities with fewer interactions~\cite{DBLP:journals/ipm/SanchezB21}. Similarly, \citeauthor{DBLP:journals/kais/ZhangW16} use clustering over city regions to support recommendations in unfamiliar cities~\cite{DBLP:journals/kais/ZhangW16}.

Conversational recommender systems are also particularly useful in the context of travel and tourism~\cite{DBLP:journals/csur/JannachMCC21}. For example, \citeauthor{Nguyen018} proposed a novel approach to address the group POI recommendation problem~\cite{Nguyen018}. %
Their mobile system tracks and exploits users’ interactions during group discussions to recommend POIs and offer other helpful suggestions to guide the group to reach consensus.
\citeauthor{DBLP:conf/dasfaa/LiHZZLS21} present a reinforcement learning approach that integrates geographical patterns with dialogue interactions for next-POI recommendation~\cite{DBLP:conf/dasfaa/LiHZZLS21}. \citeauthor{DBLP:journals/tnn/ZhangSZWX23} propose a conversational translation approach for the next-POI recommendation problem under uncertain check-ins~\cite{DBLP:journals/tnn/ZhangSZWX23}. Their method combines a recommender module that captures both sequential information and recent preferences from conversations, with a conversational module to improve recommendation quality while minimizing conversational turns.

Another major research line has adopted operations research techniques to solve optimization problems in itinerary generation and route planning. \citeauthor{Herzog2019} give an overview of the different problem formulations~\cite{Herzog2019}, including the ``Time Dependent Team Orienteering Problem with Time Windows''~\cite{DBLP:journals/cor/GavalasKMPV15} to model route planning for travelers who wish to visit various POIs using public transport, %
the ``Vacation Planning Problem''~\cite{DBLP:journals/cor/VathisKPG23}, in which a traveler wants to explore an extensive geographical area, %
or the 
``Tourist Trip Design Problem with Travel Instructions'' integrating public transport data~\cite{DBLP:conf/recsys/AyalaGAFML17}. %

\subsection{Evaluation Practices}
\label{ss:Evaluation}

The evaluation of POI RSs has traditionally relied on offline configurations using the datasets mentioned above in \Cref{ss:Data}. Typically, the dataset is split into training and test sets, either temporally or at random, sometimes also creating a validation set.
The split can be done per user, based on each user's check-in history, or globally, treating all interactions collectively. Models are trained on the training set and evaluated on their ability to predict the venues users visit in the test set. Common evaluation metrics include Precision, Recall, nDCG, and Hit Ratio, usually measured at a fixed cutoff.
Due to the often sequential nature of the recommended POIs, some authors have proposed metrics that consider the order of visited POIs, like the pairs-F$_1$ proposed by \citeauthor{DBLP:conf/cikm/ChenOX16}~\cite{DBLP:conf/cikm/ChenOX16} or sequential adaptations by \citeauthor{DBLP:journals/umuai/SanchezB20} of the previously mentioned metrics~\cite{DBLP:journals/umuai/SanchezB20}. 

However, although practical and reproducible, this type of evaluation does not fully reflect the dynamic and interactive nature of real-world user behavior, nor does it capture user feedback or intent over time. To address the inherent limitations of offline evaluation, user studies have also been adopted to gain a better understanding of user perception, satisfaction, and behavior. For example, \citeauthor{HazwaniLIREB24} show in their user study how to mitigate the cold-start problem by showing that people who participate in storytelling experiences are more willing to provide explicit preferences~\cite{HazwaniLIREB24}. 
In addition, \citeauthor{HofschenM023} analyzed the divergence between the expected utility of users before visiting a POI and their experience after the visit, collected through a web-based interactive survey. The results showed that these two types of utility often lead to different preference patterns~\cite{HofschenM023}. 

Other studies focus on the perceived usefulness and novelty of recommendations. For example, some studies found that although algorithms can generate highly novel recommendations, many users find it difficult to appreciate items that are too unfamiliar~\cite{MassimoR21,RicciMA21}. 
Complementing these findings, research on group decision-making in tourism has explored how group preferences evolve during collaborative destination planning~\cite{DBLP:journals/jitt/DelicNNR18}. The findings highlight the need for systems that support dynamic group interactions and help build consensus.

Beyond user studies, simulation-based evaluations are also gaining traction as a way to assess recommender systems in scenarios that are difficult to observe directly, such as the tool proposed by \citeauthor{PiliponyteMR24} that allows 
DMOs to simulate the impact of an online promotion campaign~\cite{PiliponyteMR24}. This simulation uses recommendation techniques to select which destinations are promoted to each tourist. 
Simulations of interactions in a multistakeholder RS can be optimized to benefit both users and destinations by maximizing tourist usefulness, while fostering unpopular venues~\cite{Merinov024}.

\section{Pitfalls in Widely Adopted Datasets}
\label{s:data_issues}

In this section, we examine key challenges when working with data for POI recommender systems. %
Our analysis is organized by data source, where the two most prevalent primary sources in the current research, location-based social networks and user studies, are discussed in detail. We also briefly touch on more unconventional or emerging data sources that have been, or could be, used to support POI recommendations.

\subsection{Issues with Location-Based Social Network Datasets}

Location-based social network datasets, discussed in 
\Cref{ss:Data}, present several challenges that must be considered when used to develop POI recommender systems. As LBSN data is the most commonly used source for offline evaluation of POI recommendation models, we analyze it exhaustively.

\pitfall{Outdated data leads to mismatches with today's reality}{pf:outdated_data}

  Many publicly available LBSN datasets are outdated, as some of them were collected more than a decade ago~\cite{DBLP:journals/geoinformatica/0003ZWM15,DBLP:journals/csur/SanchezB22}. This temporal gap presents significant challenges: POIs may have closed, relocated, or experienced substantial changes. As a result, models trained on such data may end up recommending venues that no longer exist or are no longer relevant, undermining the system's real-world applicability of the system.
  Moreover, user behavior recorded in these datasets may no longer represent actual mobility trends, particularly in urban environments where user preferences change frequently.
  Critically, to the best of our knowledge, no publicly available POI recommendation dataset has been collected and released following the COVID-19 pandemic. This absence presents a significant gap, as the pandemic has had a lasting impact on urban mobility~\cite{Santana2023COVID,Xu2023Urban}.
  
\pitfall{Mismatches between training data and target RSs users impede reliability of model performance}{pf:train_data_mismatch}
  Several studies have revealed a mismatch between the training data used in POI recommender systems and the diverse behaviors of their target users~\cite{DBLP:journals/ipm/SanchezB21,DBLP:conf/um/SanchezD22}. For example, research on the Foursquare global-scale check-in dataset has shown that most users exhibit highly localized behavior, with check-ins concentrated in a single city~\cite{DBLP:journals/ipm/SanchezB21}. This bias limits the dataset's utility for modeling tourist activity, which often implies cross-city or international travel.
  However, reducing users to a binary classification of ``locals'' and ``tourists'' oversimplifies the complexity of real-world behavior. Each group can be further segmented based on the patterns available in their check-ins \cite{Dietz2020}. Among locals, distinctions emerge based on the types of venues they visit, while tourist behavior may vary significantly depending on the scope of their travel, i.e., if they are regional, domestic, or even intercontinental visitors, affecting the performance of different recommendation models~\cite{DBLP:conf/um/SanchezD22}. 
  A common pitfall in this context is that POI recommender systems designed for travelers are trained and evaluated with interactions provided predominantly by locals, resulting in unreliable results.
  
  Regarding temporal aspects, \citeauthor{DBLP:conf/icwsm/NoulasSMP11} analyzed 12 million Foursquare check-ins, finding a significant disparity in check-in patterns between weekdays and weekends. However, in both cases, check-ins at residential venues increased constantly throughout the day, suggesting a behavior more typical of residents than of tourists~\cite{DBLP:conf/icwsm/NoulasSMP11}. Based on our experience, researchers who consider check-in data as temporally homogeneous, not accounting for variations between weekdays and weekends, introduce additional noise.
  
  Finally, there is concern that LBSN data may not accurately represent ground truth. For example, a study of the Murcia region in Spain~\cite{DBLP:journals/gis/AgryzkovMTV17} revealed discrepancies between venues registered in Foursquare and those detected on-the-ground observations. Their study shows divergences of more than 80\% in the retail sector and more than 30\% in food and entertainment venues.  %
  Another aspect of unreliable data is the posts/check-ins performed by bots, which should be detected and excluded from the analysis~\cite{Orabi2020Detection}.

\pitfall{Incomplete data increases sparsity}{pf:incomplete_data}

  The User-POI interaction matrices in LBSNs are extremely sparse. Most users check in at only a few venues, while many POIs receive few or no check-ins. This imbalance makes it difficult to learn meaningful user preferences and intensifies the cold-start problem for both users and locations~\cite{DBLP:journals/geoinformatica/0003ZWM15}.
  To illustrate the extent of this sparsity, standard benchmarks in traditional recommendation tasks such as the Netflix and MovieLens20M datasets have interaction densities slightly above 1.5\% and 0.5\%, respectively. In contrast, LBSNs datasets like Foursquare and Gowalla have densities as low as 0.003\% and 0.005\%~\cite{DBLP:journals/csur/SanchezB22}.
  Furthermore, LBSN data is inherently incomplete as it is based on voluntary contributions from users. Unlike domains such as music or video streaming, where every user interaction is automatically logged by the system, LBSNs depend on users actively recording their visits, such as checking in or posting content.
  This self-selection introduces important bias and gaps in coverage. To address this problem, it may be beneficial to enhance existing datasets with additional contextual information using cross-domain techniques. For example, an increase in visitor activity at a particular place may be due to external factors, such as a major event (e.g., a sports match, concert, or cultural festival), while favorable weather conditions may also lead to increased mobility compared to rainy days.

\pitfall{The types of visited venues lead to bias in the datasets}{pf:type_of_venues}

  Importantly, missing data are not missing at random (MNAR). This means that the probability of missing data depends on hidden causes, such as user preferences, behaviors, platform design, and the type of users who use these platforms. This non-randomness introduces different types of biases, which can significantly affect the performance and fairness of the RS~\cite{DBLP:conf/icml/WangZ0Q19}. For instance, \citeauthor{DBLP:conf/icwsm/WangSZZ16} observed substantial discrepancies between the check-ins recorded on Foursquare and the actual mobility of the users, due to privacy considerations, lack of interest, or simply forgetting to check in in the venues~\cite{DBLP:conf/icwsm/WangSZZ16}.
  Some POIs, especially those that are considered socially sensitive or niche (e.g., adult entertainment locations or particular nightlife spots), may be underrepresented in LBSN datasets. This scarcity of check-ins is not necessarily due to the absence of visitors but rather to users' reticence to publicly associate themselves with such places. Similar behavior has been observed in other domains, such as when users omit certain music tracks from their listening history to maintain a curated public image~\cite{DROTT2018Music}.
  These types of bias, introduced by oversharing certain content or self-censorship, raise serious limitations to developing accurate and equitable POI recommender systems. When the underlying data is not only incomplete but also systematically biased, models may not capture the real preferences and behaviors of users. Discrepancies can also arise from both extraneous check-ins (e.g., users checking only to earn rewards~\cite{DBLP:conf/icwsm/WangSZZ16}) and missing check-ins at sensitive locations such as healthcare facilities or political venues~\cite{DBLP:journals/gis/AlrayesAET20}.

\pitfall{Platform-specific interactions bias temporal and spatial patterns of user behavior}{pf:platform_biases}

Online platforms exhibit different interaction patterns, which can alter how user behavior is observed and interpreted. For example, review-based platforms may offer rich information on POIs~\cite{Yan2023Personalized}, but often lack temporal accuracy. This is because reviews are typically posted some time after the original visit, hiding when the interaction actually occurred.

Similarly, research that relies on photo-sharing platforms like Flickr often uses geotagged images to infer user trajectories. In such cases, researchers attempted to map geotagged photos to real-world POIs to reconstruct users’ travel paths%
~\cite{DBLP:journals/kais/LimCLK18,Dietz2020}. Like reviews, there are challenges in the geographical or temporal inaccuracy: more than one POI may exist at the recorded coordinates (because of a low resolution or granularity, or simply because several venues exist very close to each other), users may upload content long after it is captured, and many of the available datasets from the pre-smartphone era have unreliable timestamps of creation~\cite{DBLP:journals/cacm/ThomeeSFENPBL16}. %
Due to these uncertainties, it becomes difficult to accurately reconstruct tourist trajectories.
Furthermore, geotags are not always automatically recorded but are often manually added, which introduces further uncertainty in location data. Finally, the content shared on social networks is often event-driven~\cite{Kim2021Event}. Users are more likely to post photos during festivals, holidays, or peak tourist seasons, which incorporates temporal patterns. This seasonal bias makes it difficult to extrapolate findings to less frequent periods or to formulate year-round generalizations about tourist behavior.

\pitfall{Prevalent data collection practices raise privacy and ethical issues}{pf:data_ethics}

The data collection methods for many of the datasets present additional ethical issues.
The use of LBSN data without explicit user consent, combined with %
the ability to re-identify individuals based on location patterns~\cite{DBLP:conf/icwsm/0004WSM15}, is incompatible with GDPR legislation~\cite{EuropeanParliament2016Regulation}.
Many of the freely available datasets were collected prior to entry within the GDPR, and contain sensitive and difficult to anonymize location information about users who probably did not agree to their data being processed~\cite{tonetto2024ethicalprivacyconsiderationslocation}.

\pitfall{Limited user context reduces personalization}{pf:limited_context}

LBSNs datasets often do not contain rich contextual and demographic details, such as users' age, gender, travel purpose, or personal preferences. This lack of information limits the ability to generate personalized POI recommendations that take into account important distinctions, such as whether a user is a tourist or a local, or whether a traveler is visiting a location for leisure, business, or other reasons.
While some research has explored alternative data sources to enrich the user's context in terms of weather~\cite{Trattner2018Investigating} or by analyzing user-generated photos to infer travel context~\cite{DBLP:conf/wacv/SoummPD23}, such approaches are not commonly integrated into traditional LBSN platforms.

Moreover, the influence of group dynamics is frequently neglected. It is often unclear whether a registered check-in represents individual patterns or is influenced by group contexts, such as traveling with friends, family, or colleagues. A single user may show different behavior depending on the social environment, leading to heterogeneity in activity patterns~\cite{Wang21102024}.

\subsection{Issues with Data from User Studies}
User studies are commonly used during the development and evaluation of recommender systems. In the context of recommender system research, user studies serve mainly two main purposes:
Most of these studies focus on comparing different recommendation strategies or user interfaces. In such cases, interaction data is collected under controlled conditions, specifically designed to evaluate the effects of alternative methods or models. While useful for assessing performance differences, this limits the generalization of this interaction data due to the artificial nature of the task and the experimental environment.
A smaller number of studies collected observational data by capturing user behavior or preferences without the influence of a deployed recommender system~\cite{DBLP:conf/recsys/DietzS0Q24,DBLP:journals/jitt/DelicNNR18,HofschenM023}. 
These studies are rare but valuable, as they offer insight into users' natural decision-making processes.
However, even these observational studies typically highlight only specific facets of tourist behavior or the capabilities of recommender systems. In the following, we summarize the main limitations of datasets derived from user studies.

\pitfall{Limited scalability impedes comparability with offline studies}{pf:study_scalability}
User studies that evaluate research prototypes often have a limited number of participants, raising concerns about the scalability and generalization of the findings. Typical sample sizes are of the magnitude of 100-200 participants \cite{DBLP:conf/rectour/Massimo019,DBLP:conf/rectour/BalakrishnanW21,DBLP:conf/recsys/DietzS0Q24,DBLP:journals/jitt/DelicNNR18}. Although these studies provide useful information, their small sample sizes limit the statistical power and robustness to train modern recommender models developed with respect to large-scale datasets.

\pitfall{Sampling bias skews participant demographics}{pf:sampling_bias}
While social media data introduces its own set of biases, user studies are often even more demographically skewed due to the challenges of participant recruitment. In many cases, detailed demographic information is insufficient, and there is a well-documented bias in computer science research toward recruiting WEIRD participants (Western, Educated, Industrialized, Rich, and Democratic) due to their accessibility to researchers~\cite{Linxen2021Weird,Septiandri2023WEIRD}.
This overrepresentation limits the generalizability of the findings and raises concerns about the external validity of user studies in POI recommendation research.

\pitfall{Short-term observations limit temporal generalizability}{pf:short_term_observations}
Consistent with the limitations discussed above, user studies typically capture only short-term behavior, offering a snapshot of user interactions over a limited period. As a result, user studies are typically ill-suited for examining long-term tourism dynamics, such as evolving preferences, seasonal patterns, or repeat visit behaviors.

\section{Pitfalls in Algorithm Design}
\label{sec:alg}

As discussed in \Cref{s:algorithms}, a wide variety of algorithms have been proposed for POI recommendation. In this section, we focus on the general challenges that most algorithms tend to face in this recommendation domain.

\pitfall{Emphasis on accuracy metrics in algorithm design despite unreliable measurements}{pf:accuracy_metrics}
he incompleteness of LBSN datasets (cf. Pitfalls \ref{pf:incomplete_data} and \ref{pf:type_of_venues}) presents a significant challenge for any algorithm aiming to perform well in practice. Since recorded information consists of voluntary check-ins, POI RSs cannot capture the full scope of user behavior--as is possible in general recommendation systems, where datasets stem from interactions with a deployed recommender system~\cite{DBLP:conf/sigir/CanamaresC18}. This limited information forces models to learn only a partial view of real-world behavioral patterns, leading to unreliable outcome metrics. Moreover, many POI recommendation approaches formally optimize accuracy, despite the fact that the datasets themselves do not fully reflect user behavior and are inherently unreliable.
Hence, even when accuracy is targeted, this contributes to observing lower estimated values, compared to other RS application areas~\cite{MassimoR21}. This issue will also be discussed in the evaluation pitfalls section (\Cref{sec:eval}), but here we emphasize the importance of a more cautious selection of the target accuracy metric, which is optimized in the model.

\pitfall{Rigidity and lack of adaptability of the algorithms reduce contextual relevance}{pf:alg_rigidity}%
Often, the recommendation algorithms address a formal problem that does not match exactly the true application problem, as discussed in \Cref{sec:problem}.
A common consequence of that is the lack of a desirable property: dynamic adaptability to the evolving state of individual user intents and contextual factors. In fact, tourists' general motivations for visiting a destination---whether leisure, business, or cultural exploration--may significantly influence the preferences. Moreover, while visiting a destination, it is common for users to include a variety of types of POIs in their itinerary, depending on their context, for instance, searching for a place to relax when tired. Hence, algorithms that lack a proper mechanism to adapt general personalization techniques to incorporate dynamic contextual information do not provide effective support to users and will be perceived as static and not responsive.
To fill this gap, there is a need to develop systems that can respond to contextual changes and facilitate decision-making during the trip, not only before the travel starts. For example, in~\cite{DBLP:journals/eswa/OtakiB25}, the authors propose a framework which allows users to modify itineraries directly on a map to better adapt their personal interests, which in the long-term might only gradually shift, but can vary widely in the short run, or be very time-sensitive~\cite{DBLP:journals/ijon/LiuYXYHW22}.
Advanced models that integrate multi-objective optimization, taking into account factors like user satisfaction, diversity, and contextual relevance, have shown promise in addressing these challenges~\cite{DBLP:journals/eswa/ChenZLW23}. However, the true problem, to design an effective and unobtrusive context monitor component and to adapt recommendations to the obtained information, remains.

\pitfall{Popularity bias in recommendation algorithms compromises fairness, diversity, and sustainable tourism}{pf:algo_bias}
Popularity bias is a common problem in RS, and is even more pronounced in POI RSs. The result is that well-known venues are disproportionately recommended, eclipsing less popular but potentially interesting POIs. While popular venues are worth seeing for many tourists and recommending them increases confidence in the RS, excessively focusing on these POIs presents some problems. In fact, recommending popular POIs might even exacerbate the massification of central venues, negatively impacting the quality of tourists' experiences and the well-being of the local community. In fact, exposure bias, where certain items are presented more frequently to users regardless of their relevance, might only increase the severity of the problem~\cite{DBLP:journals/datamine/SanchezBB23}. Such biases not only limit the diversity of recommendations, but also raise concerns about fairness, as popular POIs may not be well suited to niche types of travelers~\cite{DBLP:journals/ipm/YalcinB22}. Hence, addressing these challenges requires developing algorithms that can mitigate such biases while still providing accurate and personalized recommendations to all types of users.

\pitfall{Disregarding multistakeholder setting hinders fair and sustainable competition}{pf:multistakeholder}
Current algorithms, developed by academic research, often target the end-user, that is, the user who visits the recommended POIs, ignoring other stakeholders such as service providers or the local community. This focus can lead to unacceptable outcomes for the broader ecosystem~\cite{DBLP:journals/fdata/BanerjeeBW24}, and can even produce unintended negative consequences for the target users of the system, such as overtourism and lack of coverage of many small businesses. Some works have shown that it is possible to incorporate the interests of different multistakeholders while preserving the utility of the recommendations. \citeauthor{DBLP:conf/rectour/BalakrishnanW21} show that the end users are sensitive to changes in the priorities of the recommendations and accept the inclusion of the interests of other stakeholders, as long as there is transparency and the recommendations maintain some utility for them\cite{DBLP:conf/rectour/BalakrishnanW21}. Tuning a stakeholder's balance parameter can %
simultaneously improve user satisfaction while reducing overcrowding in popular locations~\cite{Merinov024}. That study highlights that a balanced promotion of sustainable POIs can align different stakeholder goals more effectively.

\pitfall{Lack of transparency and explainability compromises user trust and limits system adoption.}{pf:transparency}
As a multistakeholder environment, POI recommendations affect not only user decisions but also local economies. Relying on black-box models without providing any explanation to users, may cause them not to trust the system and, therefore, to stop using it. Although recent approaches based on LLMs allow for more instinctive interaction with the system using natural language \cite{banerjee2025synthtrips}, the validity and truthfulness of the recommendations are not easy to verify. Hence, since visiting a POI affects both the time and the economic cost of the user, it is a fundamental requirement to verify the explainability of the models.

\section{Pitfalls in the Evaluation}
\label{sec:eval}

In this section, we discuss important challenges that should be faced when evaluating POI recommendation approaches; some of them are direct consequences of the previously discussed issues regarding data and algorithms.

\pitfall{Incomplete measurements misalign with user goals, overlook trust and diversity}{pf:incomplete_measurements}
There is often a mismatch between the evaluation metrics used and the actual users' tasks and objectives. For example, while information retrieval 
metrics like precision or recall are adequate to measure the system's ability to retrieve POIs that match user queries, they will not effectively evaluate the system's ability to help users explore a destination's offerings or to plan a complete visit (multiple POIs).  
Besides, in the tourism domain, it is more important to support, and therefore evaluate, the ability of the system to allow the discovery of novel POI information, rather than enabling a simple information search, which focuses on the retrieval of items, based on the user's explicit queries, which are sometimes impossible to formulate for the user~\cite{DBLP:journals/expert/StaabWRZGFPK02,gretzel2011intelligent}. 

This is related to the multifaceted nature of the application domain: travelers want to capture the variety of destinations' offerings, hence they need help to reflect on, compare, and assess the recommended POIs. Therefore, POI RSs should also be evaluated on their ability to represent the diversity of a destination's attractions, not just their ability to identify individual POIs that match the user's tastes.

At the same time, a solid evaluation of a tourist RS should take into account the trust of the user towards the system, as recommendations in this domain can often be considered as advertisements, especially those present in operational systems, which are created by marketing departments. Moreover, users usually invest a large amount of money and time in configuring a holiday, so receiving poor recommendations has a significant negative impact. Therefore, it is essential that the systems must be reliable to be useful in this domain \cite{BERHANU2020e03439}.

\pitfall{Neglect of user-centric factors weakens usability, trust, and contextual relevance in real-world deployments}{pf:user_factors}
Tourist experiences are closely related to interactive systems, hence centered around human experience, where ergonomics, usability, and accessibility are needed to provide useful systems for users
\cite{DBLP:journals/jitt/StankovG20,DBLP:conf/hcitoch/Ficarra10}.
Therefore, evaluations must assess and measure these aspects. Without proper validation of these aspects, it is risky to launch a POI RS in an operational environment.
In this context, incorporating explanations and transparency in the user interactions should be another dimension to consider in the evaluation, especially when running user studies. This would increase the trust in the system and make the tourism experiences more adaptive and comprehensible \cite{DBLP:journals/es/LealVMB25}.

Moreover, important aspects for tourists, such as distance to POIs, price, and user experience (e.g., overcrowding places), are frequently omitted in both recommendation algorithms and their evaluations. Incorporating these factors into the evaluation is essential to improve user satisfaction and provide contextually relevant suggestions \cite{park2019travel}.

\pitfall{Lack of user or item segmentation leads to generic recommendations that misalign with traveler types}{pf:item_segmentation}
Distinguishing between different types of users (such as locals and tourists) or items (popular or niche) is also crucial in the evaluation stage of the system~\cite{DBLP:conf/um/SanchezD22}. 
However, most of the current research does not consider users or items based on their behavior or characteristics, leading to generic recommendations that may not align with individual user needs or other multistakeholder goals.

In particular, users' segmentation has a deep impact in dealing with the cold-start problem, which is, in fact, common to many RS application domains, where new users (or POIs) lack sufficient interaction data to be adequately served with (or in) recommendations. 
Some approaches attempt to mitigate this problem by incorporating auxiliary information or taking advantage of group behaviors~\cite{DBLP:conf/ijcai/Sun00ZC021}, but a universally effective solution is still missing. In fact, standard evaluation procedures do not explicitly consider these conditions as part of the holistic scenario where tourism RSs should continue providing \textit{good} suggestions.

\pitfall{Inconsistent evaluation strategies with limited metrics hinder fair comparisons and obscure real-world effectiveness}{pf:inconsistent_eval_strategies}
The lack of standardized data splitting methods (e.g., random vs. temporal, user-based vs. global) can lead to data leakage and unreliable performance evaluations \cite{DBLP:journals/tois/JiS0L23}. 
It is essential for studies to clearly specify their data splitting strategies and other evaluation settings
to ensure reproducibility and allow fair comparisons of alternative models~\cite{DBLP:conf/recsys/MengMMO20,DBLP:journals/csur/SanchezB22}.

At the same time, most of the current literature prioritizes ranking accuracy metrics like Precision, Recall, and nDCG, and often ignores other critical dimensions such as novelty, diversity, user satisfaction, and fairness~\cite{DBLP:journals/csur/SanchezB22, DBLP:journals/eswa/WerneckSSPMR21}. For example, algorithms trained to optimize accuracy may repeatedly suggest popular POIs, which are also liked by many tourists, but this also produces a lack of variety and ultimately user disengagement.
These limited evaluation perspectives result in poorly diversified recommendations and users who are unaware of new or less popular venues, hence providing a bias picture of how the algorithm would perform in a real, multistakeholder environment.

\pitfall{Lack of reproducibility hinders scientific progress through comparative evaluation}{pf:reproducilibity}
Although it is becoming increasingly common to publish the source code of the core RSs algorithm in public repositories, the reproducibility of the published results is still problematic. For example, \citeauthor{DBLP:journals/csur/SanchezB22} analyzed POI recommendation papers between 2011 and 2020 in \cite{DBLP:journals/csur/SanchezB22} and they observed that, out of $310$ proposals, only $13$ provided the code of their models. It should also be noted that sometimes even the bare code is not sufficient to determine that the conducted experiments are general and reproducible \cite{Dacrema2021}. 
Hence, it is necessary to ensure that the code is easily adaptable and modifiable to experiment with other datasets or under different conditions, so that the overall evaluation remains comparable and fully reproducible.

\section{A Research Agenda to Overcome the Pitfalls of Current POI Recommender Systems}
\label{sec:agenda}

As discussed in the previous sections, building more useful POI recommender systems still faces a number of diverse issues. In this section, we highlight a few research directions aimed at overcoming these issues. We present the proposed research directions as six potential solutions: 
Multistakeholder design, context-awareness, data collection, trustworthiness design, novel interactions, and real-world evaluations.

\solution{Multistakeholder Design}{sol:multistakeholder_design} As we previously mentioned, a proper definition of the POI recommendation task requires the integration of multiple stakeholders' objectives~\cite{AbdollahpouriB22}. Commercial platform owners are aware of this dimension and have already implemented recommendation techniques, which are only partially described, that balance their profit with personalization, that is, matching offers with observed tourists' preferences~\cite{Goldenberg22}.
Conversely, public organizations, i.e., destination management companies, lack the competence and the technologies to promote an effective transition from the role of tourist information providers to tourists' flow managers. In fact, nowadays, the focus and the driver of research should be destination sustainability~\cite{Mauro24}. In this sense, many destination management organizations are struggling with overtourism and are failing to enhance the development of more peripheral POIs and locations, which is a vital goal for these organizations. To help such organizations and build more useful POI recommender systems, the research should focus on systems that, while still enabling the attraction of tourists to the main destination's POIs, can also better distribute tourists in the full region (in time and space). Some research lines have already targeted this goal~\cite{Merinov024,Paromita23}, but more research should be specifically dedicated to the optimization of effective policies to more uniformly distribute tourists~\cite{VecchiaMQGB24}, crowdedness prediction techniques~\cite{VecchiaMQB23}, and collaboration platforms for the promotion of small tourism services~\cite{Neves24,Suayb23}, or nudging techniques for behavioral change~\cite{Mauro24}.

\solution{Context-Awareness}{sol:context_awareness} Notwithstanding the clear evidence of the importance of context in tourism preferences and behavior prediction, a limited usage of context has been made so far in POI recommender systems~\cite{VecchiaMQB23,Sanchez19}.
The main contextual dimensions that have been considered are season, weather~\cite{BraunhoferE0S14}, and group composition~\cite{DelicE0M24,NguyenRDB19}. However, those dimensions have often been considered independently, and the personalization model was still mainly driven by item-related preferences, extracted from observed behaviors. We need a more comprehensive view of the multiple contextual factors that influence tourists' preferences and behaviors, which can even shade the importance of specific individual preferences~\cite{AdomaviciusBTU22,Braunhofer016,BaltrunasLPR12}.

While physical context is now easy to monitor (position, time, weather) and use, the research should focus more on the analysis of the intentions and moods of the group of tourists to be recommended~\cite{DelicE0M24}. We need more research on the extraction of intent from observational data, in order to better understand, what apparently tourists are looking for, and at the same time we must develop more solutions to extract context and intent factors from the user--system interaction, i.e., by mining text content or by interacting with the user with natural language interfaces. 
To this end, we believe some recent research works are of interest; for example, real-time analysis of tourists' discussions~\cite{Karahodza2025GroupDynamics},
behavioral clustering of observed tourists' behaviors~\cite{Massimo023}, identification of the best context for a visit, or methods for detecting context importance~\cite{BaltrunasLPR12}.

\solution{Data Collection}{sol:data_collection} It has often been mentioned (and discussed in Pitfalls \ref{pf:type_of_venues} and \ref{pf:platform_biases}) that state of the art POI RSs are limited by their preference models, which are acquired on limited and biased observation logs. 
Commercial stakeholders are in a better position, as they can rely on complete logs of real service purchases. However, the scope of their recorded interactions is often limited to a single type of POI or service (e.g., accommodations or events). Destination management organizations are better positioned, having access to various data sources: transportation, hotel reservations, culture-related tickets, or information requests and dialogues with tourist information points. Marketing destination companies are starting to analyze these data sources, with the help of commercial enterprises offering CRM tools, but, at the same time, very rarely is this data shared with the academic research~\cite{VecchiaMQGB24}.
Therefore, the development and dissemination of these datasets is of primary importance, and specific actions to promote and enable this data-sharing activity must be identified. We also believe in the potential to generate synthetic data that combines real observations with debiasing and data augmentation techniques~\cite{MerinovM023,EmamgholizadehDR24}.

\solution{Trustworthiness Design}{sol:trustworthiness_design}  A major limitation of current POI recommender systems is their lack of transparency and, more broadly, trustworthiness~\cite{WangZWR24}. Visiting a POI often involves time and cost, and poor recommendations can negatively affect a tourist's experience. This helps explain why tourists tend to favor popular places and invest time in reading reviews, which signal social validation.
These simple approaches to decision-making have been shown to be able to increase the user confidence in the choice~\cite{JamesonWF22}. These goals should be faced by new types of RSs where the tourist will have a larger confidence that the recommendations are genuinely constructed to better match their needs and wants. Such systems are required by the European legislation (e.g., Digital Service Act), but the real offer of such solutions is now very limited. 
Therefore, we need algorithms and testing approaches that can ensure the multiple stakeholders that the identified recommendations optimize the multiple criteria that the stakeholders have independently defined~\cite{AbdollahpouriB22}. 
Moreover, consumers should be made aware of such a mediation and be put in the position to eventually opt out. This requires a non-trivial level of recommendation process transparency, adapted to be understandable and usable by various types of tourists.

\solution{Novel Interactions}{sol:novel_interactions}
Nowadays, tourists rely on a multitude of specialized apps, each dealing with specific aspects of a trip, such as accommodations, transport (e.g., flights, trains), restaurants, or local venues. While these tools offer specific services, no application can integrate all of them at once, hence requiring users to manually integrate information from various sources. In fact, this segmentation overlooks the interconnection of trip elements. For example, a recommended attraction may not align with the user’s hotel, the available transportation options, or the user's time constraints. These discrepancies can result in suboptimal itineraries that do not take into account the overall needs and preferences of travelers. To meet these challenges, the next generation of POI recommender systems must take a more holistic approach to trip planning. 
This involves integrating various trip components, such as transportation, accommodations, activities, and restaurants, into a coherent framework that takes into account the user’s context, preferences, and constraints. Using comprehensive data sources and advanced algorithms, these systems can generate personalized end-to-end itineraries that increase user satisfaction and streamline the trip planning process.
A technical approach to glue the output of multiple systems comes from the usage of LLMs in conversational systems~\cite{Nguyen018,DelicE0024,Camilleri23}. The future generation of RSs may be perceived as agents that, relying on multiple information and recommendation components, offer the necessary middleware to integrate these components and offer a unique access point. Some recent works on AI supporting group discussion go exactly in this direction~\cite{muca,ricci2025CAJO}.

\solution{Real-World Evaluations}{sol:real_world_evaluations}
Traditional evaluation methods for tourism RSs often rely on offline metrics (e.g., precision, recall) that are easy to measure. Although certainly useful, they only measure the system performance on the basis of historical data. They cannot assess the quality of the recommendations when they are influenced by multiple factors, such as ephemeral preferences, contextual factors, or dynamics of decision-making. To fill this gap, Living Labs have emerged as a promising approach for real-world evaluation of tourism recommender systems. Living labs are user-centered innovation ecosystems that facilitate co-creation and experimentation in real-world environments~\cite{ballon2015living}.
In living labs, diverse stakeholders (including tourists, local residents, service providers, and policy makers) can design, test, and create different solutions that benefit multiple stakeholders. The collaborative nature of living labs, thanks to the continuous feedback and adaptation, ensures that the RS remains responsible for the changing needs and preferences of the users. In general, we need to experiment with new forms of online evaluations where it could be simpler and more efficient to run targeted tests of specific system properties.

Notwithstanding the potential increase in efficiency of online tests, the need to perform offline experiments remains. We believe that the recent trend of performing calibrated simulations of tourist behaviors can enhance the ability of the offline evaluation step to better select good system candidates to test online, as new research works indicate~\cite{NguyenRDB19,Merinov024}.

\begin{table}[tb]
\caption{\textbf{Pitfalls mapped to potential solutions.} ``\ding{52}\ding{52}'' indicates the column as a primary solution to the pitfall (row), while ``\ding{51}'' represents a secondary, i.e., more indirect connection, solution.}
\label{tab:pitfalls_solutions}
\renewcommand{\arraystretch}{1.2}
\small
\begin{tabularx}{\linewidth}{lp{4cm}XXXXXX}
\toprule
\multicolumn{2}{l}{} & \multicolumn{6}{c}{\textbf{Solutions}} \\
\cmidrule(lr){3-8}
\textbf{Cat.} & \textbf{Pitfall} & Multi\-stake\-holder Design & Context-Awareness & Data\-Collection & Trust\-wor\-thi\-ness Design & Novel\-Interactions & Real-World Evaluations \\
\midrule
\multirow{7}{*}{\parbox{.88cm}{\textbf{LBSN Data}}} & \cellcolor{gray!15} (\ref{pf:outdated_data}) Outdated Data & \cellcolor{yellow!20} \ding{51} & \cellcolor{yellow!20} \ding{51} & \cellcolor{blue!35} \ding{52}\ding{52} & \cellcolor{gray!15}  & \cellcolor{yellow!20} \ding{51} & \cellcolor{yellow!20} \ding{51} \\
 & (\ref{pf:train_data_mismatch}) Data Mismatches &  &  & \cellcolor{yellow!20} \ding{51} &  &  & \cellcolor{blue!35} \ding{52}\ding{52} \\
 & \cellcolor{gray!15} (\ref{pf:incomplete_data}) Incomplete Data & \cellcolor{yellow!20} \ding{51} & \cellcolor{yellow!20} \ding{51} & \cellcolor{blue!35} \ding{52}\ding{52} & \cellcolor{gray!15}  & \cellcolor{yellow!20} \ding{51} & \cellcolor{gray!15}  \\
 & (\ref{pf:type_of_venues}) POI Category Bias & \cellcolor{blue!35} \ding{52}\ding{52} &  & \cellcolor{yellow!20} \ding{51} &  & \cellcolor{yellow!20} \ding{51} &  \\
 & \cellcolor{gray!15} (\ref{pf:platform_biases}) Platform Bias & \cellcolor{yellow!20} \ding{51} & \cellcolor{yellow!20} \ding{51} & \cellcolor{yellow!20} \ding{51} & \cellcolor{gray!15}  & \cellcolor{blue!35} \ding{52}\ding{52} & \cellcolor{gray!15}  \\
 & (\ref{pf:data_ethics}) Data Collection Ethics &  &  & \cellcolor{yellow!20} \ding{51} & \cellcolor{blue!35} \ding{52}\ding{52} &  &  \\
 & \cellcolor{gray!15} (\ref{pf:limited_context}) Limited User Context & \cellcolor{gray!15}  & \cellcolor{blue!35} \ding{52}\ding{52} & \cellcolor{yellow!20} \ding{51} & \cellcolor{yellow!20} \ding{51} & \cellcolor{yellow!20} \ding{51} & \cellcolor{gray!15}  \\
\midrule
\multirow{3}{*}{\parbox{.88cm}{\textbf{User Study Data}}} & (\ref{pf:study_scalability}) Limited Scalability &  &  & \cellcolor{yellow!20} \ding{51} &  & \cellcolor{yellow!20} \ding{51} & \cellcolor{blue!35} \ding{52}\ding{52} \\
 & \cellcolor{gray!15} (\ref{pf:sampling_bias}) Sampling Bias & \cellcolor{yellow!20} \ding{51} & \cellcolor{gray!15}  & \cellcolor{blue!35} \ding{52}\ding{52} & \cellcolor{gray!15}  & \cellcolor{yellow!20} \ding{51} & \cellcolor{yellow!20} \ding{51} \\
 & (\ref{pf:short_term_observations}) Short-term Observations &  & \cellcolor{yellow!20} \ding{51} & \cellcolor{yellow!20} \ding{51} &  &  & \cellcolor{blue!35} \ding{52}\ding{52} \\
\midrule
\multirow{5}{*}{\parbox{.88cm}{\textbf{Algo-rithms}}} & \cellcolor{gray!15} (\ref{pf:accuracy_metrics}) Accuracy Emphasis & \cellcolor{yellow!20} \ding{51} & \cellcolor{gray!15}  & \cellcolor{gray!15}  & \cellcolor{yellow!20} \ding{51} & \cellcolor{yellow!20} \ding{51} & \cellcolor{blue!35} \ding{52}\ding{52} \\
 & (\ref{pf:alg_rigidity}) Algorithmic Rigidity & \cellcolor{yellow!20} \ding{51} & \cellcolor{blue!35} \ding{52}\ding{52} &  & \cellcolor{yellow!20} \ding{51} & \cellcolor{yellow!20} \ding{51} &  \\
 & \cellcolor{gray!15} (\ref{pf:algo_bias}) Popularity Bias & \cellcolor{yellow!20} \ding{51} & \cellcolor{gray!15}  & \cellcolor{blue!35} \ding{52}\ding{52} & \cellcolor{yellow!20} \ding{51} & \cellcolor{gray!15}  & \cellcolor{yellow!20} \ding{51} \\
 & (\ref{pf:multistakeholder}) Lack of Multistakeholder Aspects & \cellcolor{blue!35} \ding{52}\ding{52} &  &  & \cellcolor{yellow!20} \ding{51} &  & \cellcolor{yellow!20} \ding{51} \\
 & \cellcolor{gray!15} (\ref{pf:transparency}) Lack of Transparency & \cellcolor{gray!15}  & \cellcolor{gray!15}  & \cellcolor{gray!15}  & \cellcolor{blue!35} \ding{52}\ding{52} & \cellcolor{yellow!20} \ding{51} & \cellcolor{yellow!20} \ding{51} \\
\midrule
\multirow{5}{*}{\parbox{.88cm}{\textbf{Eval-uation}}} & (\ref{pf:incomplete_measurements}) Incomplete Measurements & \cellcolor{yellow!20} \ding{51} &  &  & \cellcolor{blue!35} \ding{52}\ding{52} & \cellcolor{yellow!20} \ding{51} & \cellcolor{yellow!20} \ding{51} \\
 & \cellcolor{gray!15} (\ref{pf:user_factors}) Neglect of User-centric Factors & \cellcolor{gray!15}  & \cellcolor{gray!15}  & \cellcolor{gray!15}  & \cellcolor{yellow!20} \ding{51} & \cellcolor{yellow!20} \ding{51} & \cellcolor{blue!35} \ding{52}\ding{52} \\
 & (\ref{pf:item_segmentation}) Lack of User and Item Segmentation & \cellcolor{blue!35} \ding{52}\ding{52} &  & \cellcolor{yellow!20} \ding{51} &  &  & \cellcolor{yellow!20} \ding{51} \\
 & \cellcolor{gray!15} (\ref{pf:inconsistent_eval_strategies}) Inconsistent Evaluation Strategies & \cellcolor{yellow!20} \ding{51} & \cellcolor{gray!15}  & \cellcolor{yellow!20} \ding{51} & \cellcolor{gray!15}  & \cellcolor{gray!15}  & \cellcolor{blue!35} \ding{52}\ding{52} \\
 & (\ref{pf:reproducilibity}) Reproducibility Issues &  &  & \cellcolor{yellow!20} \ding{51} &  &  & \cellcolor{blue!35} \ding{52}\ding{52} \\
\bottomrule
\end{tabularx}

\end{table}

\section{Lessons Learned and Conclusions}
\label{sec:conc}

We summarize the identified relationship between pitfalls and viable solutions in \autoref{tab:pitfalls_solutions}. Here we distinguish the match between pitfalls and solutions in primary and secondary, to better separate what we believe is the primary and secondary way to approach the solution of the pitfall.

Overall, closing the gap between research and practice through Real-world Evaluations (\sref{sol:real_world_evaluations}) emerges as the most important area of improvement. It is the primary solution for six pitfalls and a secondary solution seven times.
Multistakeholder Design, Data Collection, and Novel Interactions also addressed a large number of pitfalls.
However, novel interactions are the primary solution only for \pref{pf:platform_biases} (Platform Biases). This suggests that new ways to support travelers are necessary to address a broad range of issues, even if they are not always the primary remedy.

Interestingly, almost all solutions address all the categories of pitfalls, with the exception of \sref{sol:context_awareness} (Context-Awareness), which does not resolve any evaluation-related issues.
As expected, improvements in data collection practices (\pref{sol:data_collection}) primarily target pitfalls related to datasets, but they also have indirect effects on evaluation practices. For example, one of the most serious problems in POI recommendation--\pref{pf:algo_bias} (Popularity Bias)--has its root in the nature of available data sets, and we posit that algorithmic design would be different if richer and realistic datasets were available.
For this reason, we believe that several pitfalls may require multiple solutions working in coordination to improve current practices.

Some necessary improvements fall outside the scope of our proposed research agenda. For instance, improving reproducibility depends not only on the methods and data used, but also on setting high publication standards for documenting procedures and sharing code and data~\cite{Hutson2018Artificial}. Further steps, such as choosing fair baselines and conducting reproducibility and replication studies in a dedicated track\footnote{\url{https://recsys.acm.org/recsys25/call/reproducibility}} are necessary.

To conclude, POI recommendation is a research area that has received much attention, due to its apparent simplicity, but also for its clear usefulness. However, we have claimed that the majority of the proposed solutions have failed to address the important needs and demands of the involved stakeholders. We have therefore initiated the paper with a discussion of the POI recommendation problem,
before analyzing state-of-the-art algorithms and evaluation approaches. We have then identified important characteristics and challenges that need to be addressed to produce more useful and really deployable systems.
We have structured our analysis along three axes: datasets, algorithms, and evaluation approaches. 

By means of this analysis, we have shown that while LBSNs are the main source of information for POI RSs, they are composed of sparse, outdated, and highly biased data. For this reason, we argue that it is necessary to complement this information and consider more recent, complete, and even complementary data sources, such as those related to weather, events, or blogs. Regarding algorithms, we have found evidence that most of them are focused on the optimization of user-related precision metrics, ignoring other stakeholders and dimensions, such as novelty or diversity. Moreover, current solutions lack adequate support for in-trip context-dependent re-planning. In fact, it is not enough to recommend relevant items, but also to adapt the specific user profile and consider other factors such as distance, price, and popularity of the POI, so that the user is aware of the variety of POIs that can be visited. Regarding the evaluation dimension, we have evidenced that more interaction with users must be collected to better train and test the designed solutions. 

Finally, we have proposed a structured research agenda that, starting from the identified issues, introduces important directions for future work. These potential future lines are related to multistakeholder design, context-awareness, data collection, trustworthiness, novel interactions, and real-world evaluation. In particular, we consider vital the creation of datasets that incorporate not only data from LBSNs but also information from other stakeholders and contexts. Moreover, the addressed recommendation problem must take into account the interests of all the relevant stakeholders in the tourism domain. 
Because of this variety of goals and data, newer models that take into account a wider range of destination data and produce more sustainable and trustworthy recommendations, for the benefit of the full ecosystem of the tourism economy, are a primary target for future research.

\begin{acks}
This work was in part supported by grant PID2022-139131NB-I00 funded by MCIN/AEI/10.13039/ 501100011033 and by ``ERDF A way of making Europe.''
\end{acks}

\bibliographystyle{ACM-Reference-Format}
\bibliography{bibliography}


\begin{thebibliography}{129}


\ifx \showCODEN    \undefined \def \showCODEN     #1{\unskip}     \fi
\ifx \showISBNx    \undefined \def \showISBNx     #1{\unskip}     \fi
\ifx \showISBNxiii \undefined \def \showISBNxiii  #1{\unskip}     \fi
\ifx \showISSN     \undefined \def \showISSN      #1{\unskip}     \fi
\ifx \showLCCN     \undefined \def \showLCCN      #1{\unskip}     \fi
\ifx \shownote     \undefined \def \shownote      #1{#1}          \fi
\ifx \showarticletitle \undefined \def \showarticletitle #1{#1}   \fi
\ifx \showURL      \undefined \def \showURL       {\relax}        \fi
\providecommand\bibfield[2]{#2}
\providecommand\bibinfo[2]{#2}
\providecommand\natexlab[1]{#1}
\providecommand\showeprint[2][]{arXiv:#2}

\bibitem[Abdollahpouri and Burke(2022)]%
        {AbdollahpouriB22}
\bibfield{author}{\bibinfo{person}{Himan Abdollahpouri} {and}
  \bibinfo{person}{Robin Burke}.} \bibinfo{year}{2022}\natexlab{}.
\newblock \showarticletitle{Multistakeholder Recommender Systems}.
\newblock In \bibinfo{booktitle}{\emph{Recommender Systems Handbook}},
  \bibfield{editor}{\bibinfo{person}{Francesco Ricci}, \bibinfo{person}{Lior
  Rokach}, {and} \bibinfo{person}{Bracha Shapira}} (Eds.).
  \bibinfo{publisher}{Springer}, \bibinfo{pages}{647--677}.
\newblock
\showISBNx{9781071621974}
\href{https://doi.org/10.1007/978-1-0716-2197-4_17}{doi:\nolinkurl{10.1007/978-1-0716-2197-4_17}}


\bibitem[Adomavicius et~al\mbox{.}(2022)]%
        {AdomaviciusBTU22}
\bibfield{author}{\bibinfo{person}{Gediminas Adomavicius},
  \bibinfo{person}{Konstantin Bauman}, \bibinfo{person}{Alexander Tuzhilin},
  {and} \bibinfo{person}{Moshe Unger}.} \bibinfo{year}{2022}\natexlab{}.
\newblock \showarticletitle{Context-Aware Recommender Systems: From Foundations
  to Recent Developments}.
\newblock In \bibinfo{booktitle}{\emph{Recommender Systems Handbook}},
  \bibfield{editor}{\bibinfo{person}{Francesco Ricci}, \bibinfo{person}{Lior
  Rokach}, {and} \bibinfo{person}{Bracha Shapira}} (Eds.).
  \bibinfo{publisher}{Springer}, \bibinfo{pages}{211--250}.
\newblock
\showISBNx{9781071621974}
\href{https://doi.org/10.1007/978-1-0716-2197-4_6}{doi:\nolinkurl{10.1007/978-1-0716-2197-4_6}}


\bibitem[Agryzkov et~al\mbox{.}(2017)]%
        {DBLP:journals/gis/AgryzkovMTV17}
\bibfield{author}{\bibinfo{person}{Taras Agryzkov}, \bibinfo{person}{Pablo
  Mart{\'{\i}}}, \bibinfo{person}{Leandro Tortosa}, {and}
  \bibinfo{person}{Jos{\'{e}}{-}Francisco Vicent}.}
  \bibinfo{year}{2017}\natexlab{}.
\newblock \showarticletitle{Measuring urban activities using {Foursquare} data
  and network analysis: a case study of {Murcia (Spain)}}.
\newblock \bibinfo{journal}{\emph{International Journal of Geographical
  Information Science}} \bibinfo{volume}{31}, \bibinfo{number}{1}
  (\bibinfo{date}{May} \bibinfo{year}{2017}), \bibinfo{pages}{100--121}.
\newblock
\showISSN{1362-3087}
\href{https://doi.org/10.1080/13658816.2016.1188931}{doi:\nolinkurl{10.1080/13658816.2016.1188931}}


\bibitem[Alrayes et~al\mbox{.}(2020)]%
        {DBLP:journals/gis/AlrayesAET20}
\bibfield{author}{\bibinfo{person}{Fatma~S. Alrayes}, \bibinfo{person}{Alia~I.
  Abdelmoty}, \bibinfo{person}{Waleed El{-}Geresy}, {and}
  \bibinfo{person}{George Theodorakopoulos}.} \bibinfo{year}{2020}\natexlab{}.
\newblock \showarticletitle{Modelling perceived risks to personal privacy from
  location disclosure on online social networks}.
\newblock \bibinfo{journal}{\emph{International Journal of Geographical
  Information Science}} \bibinfo{volume}{34}, \bibinfo{number}{1}
  (\bibinfo{year}{2020}), \bibinfo{pages}{150--176}.
\newblock
\href{https://doi.org/10.1080/13658816.2019.1654109}{doi:\nolinkurl{10.1080/13658816.2019.1654109}}


\bibitem[Alves et~al\mbox{.}(2023)]%
        {DBLP:journals/umuai/AlvesMSCNM23}
\bibfield{author}{\bibinfo{person}{Patr{\'{\i}}cia Alves},
  \bibinfo{person}{Helena Martins}, \bibinfo{person}{Pedro~M. Saraiva},
  \bibinfo{person}{Jo{\~{a}}o Carneiro}, \bibinfo{person}{Paulo Novais}, {and}
  \bibinfo{person}{Goreti Marreiros}.} \bibinfo{year}{2023}\natexlab{}.
\newblock \showarticletitle{Group recommender systems for tourism: how does
  personality predict preferences for attractions, travel motivations,
  preferences and concerns?}
\newblock \bibinfo{journal}{\emph{User Modeling and User-Adapted Interaction}}
  \bibinfo{volume}{33}, \bibinfo{number}{5} (\bibinfo{year}{2023}),
  \bibinfo{pages}{1141--1210}.
\newblock
\href{https://doi.org/10.1007/S11257-023-09361-2}{doi:\nolinkurl{10.1007/S11257-023-09361-2}}


\bibitem[Ayala et~al\mbox{.}(2017)]%
        {DBLP:conf/recsys/AyalaGAFML17}
\bibfield{author}{\bibinfo{person}{Victor Anthony~Arrascue Ayala},
  \bibinfo{person}{Kemal~Cagin G{\"{u}}lsen}, \bibinfo{person}{Anas Alzogbi},
  \bibinfo{person}{Michael F{\"{a}}rber}, \bibinfo{person}{Marco Mu{\~{n}}iz},
  {and} \bibinfo{person}{Georg Lausen}.} \bibinfo{year}{2017}\natexlab{}.
\newblock \showarticletitle{A Delay-Robust Touristic Plan Recommendation Using
  Real-World Public Transportation Information}. In
  \bibinfo{booktitle}{\emph{2nd RecSys Workshop on Recommenders in Tourism}}
  \emph{(\bibinfo{series}{{CEUR} Workshop Proceedings},
  Vol.~\bibinfo{volume}{1906})}. \bibinfo{publisher}{CEUR-WS.org},
  \bibinfo{pages}{9--17}.
\newblock


\bibitem[Balakrishnan and W{\"{o}}rndl(2021)]%
        {DBLP:conf/rectour/BalakrishnanW21}
\bibfield{author}{\bibinfo{person}{Gokulakrishnan Balakrishnan} {and}
  \bibinfo{person}{Wolfgang W{\"{o}}rndl}.} \bibinfo{year}{2021}\natexlab{}.
\newblock \showarticletitle{Multistakeholder Recommender Systems in Tourism}.
  In \bibinfo{booktitle}{\emph{RecSys Workshop on Recommenders in Tourism}}
  \emph{(\bibinfo{series}{{CEUR} Workshop Proceedings},
  Vol.~\bibinfo{volume}{2974})}. \bibinfo{publisher}{CEUR-WS.org},
  \bibinfo{pages}{39--53}.
\newblock


\bibitem[Ballon and Schuurman(2015)]%
        {ballon2015living}
\bibfield{author}{\bibinfo{person}{Pieter Ballon} {and}
  \bibinfo{person}{Dimitri Schuurman}.} \bibinfo{year}{2015}\natexlab{}.
\newblock \showarticletitle{Living labs: concepts, tools and cases}.
\newblock \bibinfo{journal}{\emph{Info}} \bibinfo{volume}{17},
  \bibinfo{number}{4} (\bibinfo{year}{2015}).
\newblock
\showISSN{1556-5068}
\href{https://doi.org/10.2139/ssrn.2642754}{doi:\nolinkurl{10.2139/ssrn.2642754}}


\bibitem[Baltrunas et~al\mbox{.}(2012)]%
        {BaltrunasLPR12}
\bibfield{author}{\bibinfo{person}{Linas Baltrunas}, \bibinfo{person}{Bernd
  Ludwig}, \bibinfo{person}{Stefan Peer}, {and} \bibinfo{person}{Francesco
  Ricci}.} \bibinfo{year}{2012}\natexlab{}.
\newblock \showarticletitle{Context relevance assessment and exploitation in
  mobile recommender systems}.
\newblock \bibinfo{journal}{\emph{Personal and Ubiquitous Computing}}
  \bibinfo{volume}{16}, \bibinfo{number}{5} (\bibinfo{year}{2012}),
  \bibinfo{pages}{507--526}.
\newblock
\href{https://doi.org/10.1007/S00779-011-0417-X}{doi:\nolinkurl{10.1007/S00779-011-0417-X}}


\bibitem[Banerjee et~al\mbox{.}(2024)]%
        {DBLP:journals/fdata/BanerjeeBW24}
\bibfield{author}{\bibinfo{person}{Ashmi Banerjee}, \bibinfo{person}{Paromita
  Banik}, {and} \bibinfo{person}{Wolfgang Wörndl}.}
  \bibinfo{year}{2024}\natexlab{}.
\newblock \showarticletitle{A review on individual and multistakeholder
  fairness in tourism recommender systems}.
\newblock \bibinfo{journal}{\emph{Frontiers Big Data}}  \bibinfo{volume}{6}
  (\bibinfo{date}{May} \bibinfo{year}{2024}).
\newblock
\showISSN{2624-909X}
\href{https://doi.org/10.3389/fdata.2023.1168692}{doi:\nolinkurl{10.3389/fdata.2023.1168692}}


\bibitem[Banerjee et~al\mbox{.}(2020)]%
        {Banerjee2020}
\bibfield{author}{\bibinfo{person}{Ashmi Banerjee}, \bibinfo{person}{Gourab~K.
  Patro}, \bibinfo{person}{Linus~W. Dietz}, {and} \bibinfo{person}{Abhijnan
  Chakraborty}.} \bibinfo{year}{2020}\natexlab{}.
\newblock \showarticletitle{Analyzing `Near Me' Services: Potential for
  Exposure Bias in Location-based Retrieval}. In
  \bibinfo{booktitle}{\emph{International Workshop on Fair and Interpretable
  Learning Algorithms}} \emph{(\bibinfo{series}{FILA'20})}.
  \bibinfo{publisher}{{IEEE}}.
\newblock
\href{https://doi.org/10.1109/bigdata50022.2020.9378476}{doi:\nolinkurl{10.1109/bigdata50022.2020.9378476}}


\bibitem[Banerjee et~al\mbox{.}(2025)]%
        {banerjee2025synthtrips}
\bibfield{author}{\bibinfo{person}{Ashmi Banerjee}, \bibinfo{person}{Adithi
  Satish}, \bibinfo{person}{Fitri~Nur Aisyah}, \bibinfo{person}{Wolfgang
  Wörndl}, {and} \bibinfo{person}{Yashar Deldjoo}.}
  \bibinfo{year}{2025}\natexlab{}.
\newblock \showarticletitle{SynthTRIPs: A Knowledge-Grounded Framework for
  Benchmark Query Generation for Personalized Tourism Recommenders}.
\newblock \bibinfo{journal}{\emph{arXiv preprint arXiv:2504.09277}}
  (\bibinfo{year}{2025}).
\newblock
\href{https://doi.org/10.48550/arXiv.2504.09277}{doi:\nolinkurl{10.48550/arXiv.2504.09277}}
\newblock
\shownote{In press at SIGIR'25}.


\bibitem[Banik et~al\mbox{.}(2023)]%
        {Paromita23}
\bibfield{author}{\bibinfo{person}{Paromita Banik}, \bibinfo{person}{Ashmi
  Banerjee}, {and} \bibinfo{person}{Wolfgang W\"{o}rndl}.}
  \bibinfo{year}{2023}\natexlab{}.
\newblock \showarticletitle{Understanding User Perspectives on Sustainability
  and Fairness in Tourism Recommender Systems}. In
  \bibinfo{booktitle}{\emph{31st ACM Conference on User Modeling, Adaptation
  and Personalization}}. \bibinfo{publisher}{ACM}, \bibinfo{address}{New York,
  NY, USA}, \bibinfo{pages}{241--248}.
\newblock
\showISBNx{9781450398916}
\href{https://doi.org/10.1145/3563359.3597442}{doi:\nolinkurl{10.1145/3563359.3597442}}


\bibitem[Bao et~al\mbox{.}(2015)]%
        {DBLP:journals/geoinformatica/0003ZWM15}
\bibfield{author}{\bibinfo{person}{Jie Bao}, \bibinfo{person}{Yu Zheng},
  \bibinfo{person}{David Wilkie}, {and} \bibinfo{person}{Mohamed Mokbel}.}
  \bibinfo{year}{2015}\natexlab{}.
\newblock \showarticletitle{Recommendations in location-based social networks:
  a survey}.
\newblock \bibinfo{journal}{\emph{{GeoInformatica}}} \bibinfo{volume}{19},
  \bibinfo{number}{3} (\bibinfo{date}{Feb.} \bibinfo{year}{2015}),
  \bibinfo{pages}{525--565}.
\newblock
\showISSN{1384-6175}
\href{https://doi.org/10.1007/s10707-014-0220-8}{doi:\nolinkurl{10.1007/s10707-014-0220-8}}


\bibitem[Berhanu and Raj(2020)]%
        {BERHANU2020e03439}
\bibfield{author}{\bibinfo{person}{Kassegn Berhanu} {and}
  \bibinfo{person}{Sahil Raj}.} \bibinfo{year}{2020}\natexlab{}.
\newblock \showarticletitle{The trustworthiness of travel and tourism
  information sources of social media: perspectives of international tourists
  visiting Ethiopia}.
\newblock \bibinfo{journal}{\emph{Heliyon}} \bibinfo{volume}{6},
  \bibinfo{number}{3} (\bibinfo{year}{2020}), \bibinfo{pages}{e03439}.
\newblock
\showISSN{2405-8440}
\href{https://doi.org/10.1016/j.heliyon.2020.e03439}{doi:\nolinkurl{10.1016/j.heliyon.2020.e03439}}


\bibitem[Boom et~al\mbox{.}(2021)]%
        {Boom16022021}
\bibfield{author}{\bibinfo{person}{Samantha Boom}, \bibinfo{person}{Jelmer
  Weijschede}, \bibinfo{person}{Frans Melissen}, \bibinfo{person}{Ko Koens},
  {and} \bibinfo{person}{Igor~Mayer and}.} \bibinfo{year}{2021}\natexlab{}.
\newblock \showarticletitle{Identifying stakeholder perspectives and worldviews
  on sustainable urban tourism development using a Q-sort methodology}.
\newblock \bibinfo{journal}{\emph{Current Issues in Tourism}}
  \bibinfo{volume}{24}, \bibinfo{number}{4} (\bibinfo{year}{2021}),
  \bibinfo{pages}{520--535}.
\newblock
\href{https://doi.org/10.1080/13683500.2020.1722076}{doi:\nolinkurl{10.1080/13683500.2020.1722076}}


\bibitem[Braunhofer et~al\mbox{.}(2014)]%
        {BraunhoferE0S14}
\bibfield{author}{\bibinfo{person}{Matthias Braunhofer}, \bibinfo{person}{Mehdi
  Elahi}, \bibinfo{person}{Francesco Ricci}, {and} \bibinfo{person}{Thomas
  Schievenin}.} \bibinfo{year}{2014}\natexlab{}.
\newblock \showarticletitle{Context-Aware Points of Interest Suggestion with
  Dynamic Weather Data Management}. In \bibinfo{booktitle}{\emph{Information
  and Communication Technologies in Tourism 2014}} (2013).
  \bibinfo{publisher}{Springer}, \bibinfo{pages}{87--100}.
\newblock
\showISBNx{9783319039732}
\href{https://doi.org/10.1007/978-3-319-03973-2_7}{doi:\nolinkurl{10.1007/978-3-319-03973-2_7}}


\bibitem[Braunhofer and Ricci(2016)]%
        {Braunhofer016}
\bibfield{author}{\bibinfo{person}{Matthias Braunhofer} {and}
  \bibinfo{person}{Francesco Ricci}.} \bibinfo{year}{2016}\natexlab{}.
\newblock \showarticletitle{Contextual Information Elicitation in Travel
  Recommender Systems}. In \bibinfo{booktitle}{\emph{Information and
  Communication Technologies in Tourism 2016}} (2016).
  \bibinfo{publisher}{Springer}, \bibinfo{pages}{579--592}.
\newblock
\showISBNx{9783319282312}
\href{https://doi.org/10.1007/978-3-319-28231-2_42}{doi:\nolinkurl{10.1007/978-3-319-28231-2_42}}


\bibitem[Brilhante et~al\mbox{.}(2013)]%
        {DBLP:conf/cikm/BrilhanteMNPR13}
\bibfield{author}{\bibinfo{person}{Igo~Ramalho Brilhante},
  \bibinfo{person}{Jos{\'{e}} Ant{\^{o}}nio~Fernandes de Mac{\^{e}}do},
  \bibinfo{person}{Franco~Maria Nardini}, \bibinfo{person}{Raffaele Perego},
  {and} \bibinfo{person}{Chiara Renso}.} \bibinfo{year}{2013}\natexlab{}.
\newblock \showarticletitle{Where shall we go today?: planning touristic tours
  with tripbuilder}. In \bibinfo{booktitle}{\emph{22nd {ACM} International
  Conference on Information \& Knowledge Management}}
  \emph{(\bibinfo{series}{CIKM'13})}. \bibinfo{publisher}{{ACM}},
  \bibinfo{pages}{757--762}.
\newblock
\href{https://doi.org/10.1145/2505515.2505643}{doi:\nolinkurl{10.1145/2505515.2505643}}


\bibitem[Burke and Ramezani(2011)]%
        {Burke2011}
\bibfield{author}{\bibinfo{person}{Robin~D. Burke} {and}
  \bibinfo{person}{Maryam Ramezani}.} \bibinfo{year}{2011}\natexlab{}.
\newblock \showarticletitle{Recommender Systems Handbook}.
\newblock \bibinfo{publisher}{Springer}, \bibinfo{address}{Boston, MA, USA},
  Chapter Matching Recommendation Technologies and Domains,
  \bibinfo{pages}{367--386}.
\newblock
\showISBNx{978-0-387-85820-3}
\href{https://doi.org/10.1007/978-0-387-85820-3_11}{doi:\nolinkurl{10.1007/978-0-387-85820-3_11}}


\bibitem[Camilleri and Troise(2023)]%
        {Camilleri23}
\bibfield{author}{\bibinfo{person}{Mark~Anthony Camilleri} {and}
  \bibinfo{person}{Ciro Troise}.} \bibinfo{year}{2023}\natexlab{}.
\newblock \showarticletitle{Chatbot recommender systems in tourism: A
  systematic review and a benefit-cost analysis}. In
  \bibinfo{booktitle}{\emph{8th International Conference on Machine Learning
  Technologies}}. \bibinfo{publisher}{ACM}, \bibinfo{address}{New York, NY,
  USA}, \bibinfo{pages}{151--156}.
\newblock
\showISBNx{9781450398329}
\href{https://doi.org/10.1145/3589883.3589906}{doi:\nolinkurl{10.1145/3589883.3589906}}


\bibitem[Ca{\~{n}}amares and Castells(2018)]%
        {DBLP:conf/sigir/CanamaresC18}
\bibfield{author}{\bibinfo{person}{Roc{\'{\i}}o Ca{\~{n}}amares} {and}
  \bibinfo{person}{Pablo Castells}.} \bibinfo{year}{2018}\natexlab{}.
\newblock \showarticletitle{Should {I} Follow the Crowd?: {A} Probabilistic
  Analysis of the Effectiveness of Popularity in Recommender Systems}. In
  \bibinfo{booktitle}{\emph{41st International {ACM} {SIGIR} Conference on
  Research \& Development in Information Retrieval}}.
  \bibinfo{publisher}{{ACM}}, \bibinfo{pages}{415--424}.
\newblock
\href{https://doi.org/10.1145/3209978.3210014}{doi:\nolinkurl{10.1145/3209978.3210014}}


\bibitem[Chen et~al\mbox{.}(2016)]%
        {DBLP:conf/cikm/ChenOX16}
\bibfield{author}{\bibinfo{person}{Dawei Chen}, \bibinfo{person}{Cheng~Soon
  Ong}, {and} \bibinfo{person}{Lexing Xie}.} \bibinfo{year}{2016}\natexlab{}.
\newblock \showarticletitle{Learning Points and Routes to Recommend
  Trajectories}. In \bibinfo{booktitle}{\emph{25th ACM International on
  Conference on Information and Knowledge Management}}
  \emph{(\bibinfo{series}{CIKM'16})}. \bibinfo{publisher}{{ACM}},
  \bibinfo{address}{New York, NY, USA}, \bibinfo{pages}{2227--2232}.
\newblock
\showISBNx{9781450340731}
\href{https://doi.org/10.1145/2983323.2983672}{doi:\nolinkurl{10.1145/2983323.2983672}}


\bibitem[Chen et~al\mbox{.}(2023)]%
        {DBLP:journals/eswa/ChenZLW23}
\bibfield{author}{\bibinfo{person}{Lei Chen}, \bibinfo{person}{Guixiang Zhu},
  \bibinfo{person}{Weichao Liang}, {and} \bibinfo{person}{Youquan Wang}.}
  \bibinfo{year}{2023}\natexlab{}.
\newblock \showarticletitle{Multi-objective reinforcement learning approach for
  trip recommendation}.
\newblock \bibinfo{journal}{\emph{Expert Systems with Applications}}
  \bibinfo{volume}{226} (\bibinfo{year}{2023}), \bibinfo{pages}{120145}.
\newblock
\href{https://doi.org/10.1016/J.ESWA.2023.120145}{doi:\nolinkurl{10.1016/J.ESWA.2023.120145}}


\bibitem[Dacrema et~al\mbox{.}(2021)]%
        {Dacrema2021}
\bibfield{author}{\bibinfo{person}{Maurizio~Ferrari Dacrema},
  \bibinfo{person}{Simone Boglio}, \bibinfo{person}{Paolo Cremonesi}, {and}
  \bibinfo{person}{Dietmar Jannach}.} \bibinfo{year}{2021}\natexlab{}.
\newblock \showarticletitle{A Troubling Analysis of Reproducibility and
  Progress in Recommender Systems Research}.
\newblock \bibinfo{journal}{\emph{{ACM} Transactions on Information Systems}}
  \bibinfo{volume}{39}, \bibinfo{number}{2} (\bibinfo{date}{April}
  \bibinfo{year}{2021}), \bibinfo{pages}{1--49}.
\newblock
\href{https://doi.org/10.1145/3434185}{doi:\nolinkurl{10.1145/3434185}}


\bibitem[Delic et~al\mbox{.}(2024a)]%
        {DelicE0024}
\bibfield{author}{\bibinfo{person}{Amra Delic}, \bibinfo{person}{Hanif
  Emamgholizadeh}, \bibinfo{person}{Thuy~Ngoc Nguyen}, {and}
  \bibinfo{person}{Francesco Ricci}.} \bibinfo{year}{2024}\natexlab{a}.
\newblock \showarticletitle{{CHARM:} a Group Decision Making Support Chatbot}.
  In \bibinfo{booktitle}{\emph{29th International Conference on Intelligent
  User Interfaces}}. \bibinfo{publisher}{{ACM}}, \bibinfo{pages}{7--10}.
\newblock
\href{https://doi.org/10.1145/3640544.3645220}{doi:\nolinkurl{10.1145/3640544.3645220}}


\bibitem[Delic et~al\mbox{.}(2024b)]%
        {DelicE0M24}
\bibfield{author}{\bibinfo{person}{Amra Delic}, \bibinfo{person}{Hanif
  Emamgholizadeh}, \bibinfo{person}{Francesco Ricci}, {and}
  \bibinfo{person}{Judith Masthoff}.} \bibinfo{year}{2024}\natexlab{b}.
\newblock \showarticletitle{Supporting Group Decision-Making: Insights from a
  Focus Group Study}. In \bibinfo{booktitle}{\emph{32nd {ACM} Conference on
  User Modeling, Adaptation and Personalization}}. \bibinfo{publisher}{{ACM}},
  \bibinfo{pages}{301--306}.
\newblock
\href{https://doi.org/10.1145/3627043.3659538}{doi:\nolinkurl{10.1145/3627043.3659538}}


\bibitem[Delic et~al\mbox{.}(2018)]%
        {DBLP:journals/jitt/DelicNNR18}
\bibfield{author}{\bibinfo{person}{Amra Delic}, \bibinfo{person}{Julia
  Neidhardt}, \bibinfo{person}{Thuy~Ngoc Nguyen}, {and}
  \bibinfo{person}{Francesco Ricci}.} \bibinfo{year}{2018}\natexlab{}.
\newblock \showarticletitle{An observational user study for group recommender
  systems in the tourism domain}.
\newblock \bibinfo{journal}{\emph{Information Technology \& Tourism}}
  \bibinfo{volume}{19}, \bibinfo{number}{1-4} (\bibinfo{year}{2018}),
  \bibinfo{pages}{87--116}.
\newblock
\href{https://doi.org/10.1007/S40558-018-0106-Y}{doi:\nolinkurl{10.1007/S40558-018-0106-Y}}


\bibitem[Dietz et~al\mbox{.}(2024)]%
        {DBLP:conf/recsys/DietzS0Q24}
\bibfield{author}{\bibinfo{person}{Linus~W. Dietz}, \bibinfo{person}{Sanja
  Scepanovic}, \bibinfo{person}{Ke Zhou}, {and} \bibinfo{person}{Daniele
  Quercia}.} \bibinfo{year}{2024}\natexlab{}.
\newblock \showarticletitle{Exploratory Analysis of Recommending Urban Parks
  for Health-Promoting Activities}. In \bibinfo{booktitle}{\emph{18th {ACM}
  Conference on Recommender Systems}}. \bibinfo{publisher}{{ACM}},
  \bibinfo{pages}{1131--1135}.
\newblock
\href{https://doi.org/10.1145/3640457.3691712}{doi:\nolinkurl{10.1145/3640457.3691712}}


\bibitem[Dietz et~al\mbox{.}(2020)]%
        {Dietz2020}
\bibfield{author}{\bibinfo{person}{Linus~W. Dietz}, \bibinfo{person}{Avradip
  Sen}, \bibinfo{person}{Rinita Roy}, {and} \bibinfo{person}{Wolfgang
  W\"orndl}.} \bibinfo{year}{2020}\natexlab{}.
\newblock \showarticletitle{Mining Trips from Location-based Social Networks
  for Clustering Travelers and Destinations}.
\newblock \bibinfo{journal}{\emph{Information Technology {\&} Tourism}}
  \bibinfo{volume}{22}, \bibinfo{number}{1} (\bibinfo{date}{March}
  \bibinfo{year}{2020}), \bibinfo{pages}{131--166}.
\newblock
\showISSN{1098-3058}
\href{https://doi.org/10.1007/s40558-020-00170-6}{doi:\nolinkurl{10.1007/s40558-020-00170-6}}


\bibitem[Dietz et~al\mbox{.}(2022)]%
        {Dietz2022a}
\bibfield{author}{\bibinfo{person}{Linus~W. Dietz}, \bibinfo{person}{Mete
  Sertkan}, \bibinfo{person}{Saadi Myftija}, \bibinfo{person}{Sameera~Thimbiri
  Palage}, \bibinfo{person}{Julia Neidhardt}, {and} \bibinfo{person}{Wolfgang
  Wörndl}.} \bibinfo{year}{2022}\natexlab{}.
\newblock \showarticletitle{A Comparative Study of Data-driven Models for
  Travel Destination Characterization}.
\newblock \bibinfo{journal}{\emph{Frontiers in Big Data}}  \bibinfo{volume}{5}
  (\bibinfo{date}{April} \bibinfo{year}{2022}).
\newblock
\showISSN{2624-909X}
\href{https://doi.org/10.3389/fdata.2022.829939}{doi:\nolinkurl{10.3389/fdata.2022.829939}}


\bibitem[Drott(2018)]%
        {DROTT2018Music}
\bibfield{author}{\bibinfo{person}{Eric~A. Drott}.}
  \bibinfo{year}{2018}\natexlab{}.
\newblock \showarticletitle{Music as a Technology of Surveillance}.
\newblock \bibinfo{journal}{\emph{Journal of the Society for American Music}}
  \bibinfo{volume}{12}, \bibinfo{number}{3} (\bibinfo{date}{July}
  \bibinfo{year}{2018}), \bibinfo{pages}{233--267}.
\newblock
\showISSN{1752-1971}
\href{https://doi.org/10.1017/s1752196318000196}{doi:\nolinkurl{10.1017/s1752196318000196}}


\bibitem[Emamgholizadeh et~al\mbox{.}(2024)]%
        {EmamgholizadehDR24}
\bibfield{author}{\bibinfo{person}{Hanif Emamgholizadeh}, \bibinfo{person}{Amra
  Delic}, {and} \bibinfo{person}{Francesco Ricci}.}
  \bibinfo{year}{2024}\natexlab{}.
\newblock \showarticletitle{Predicting Group Choices from Group Profiles}.
\newblock \bibinfo{journal}{\emph{ACM Transactions on Interactive Intelligent
  Systems}} \bibinfo{volume}{14}, \bibinfo{number}{1} (\bibinfo{year}{2024}),
  \bibinfo{pages}{7:1--7:27}.
\newblock
\href{https://doi.org/10.1145/3639710}{doi:\nolinkurl{10.1145/3639710}}


\bibitem[{European Parliament} and {Council of the European Union}(2016)]%
        {EuropeanParliament2016Regulation}
\bibfield{author}{\bibinfo{person}{{European Parliament}} {and}
  \bibinfo{person}{{Council of the European Union}}.}
  \bibinfo{year}{2016}\natexlab{}.
\newblock \bibinfo{booktitle}{\emph{Regulation ({EU}) 2016/679 of the
  {European} {Parliament} and of the {Council}}}.
\newblock OJ L 119, 4.5.2016, p. 1--88.
\newblock
\urldef\tempurl%
\url{https://data.europa.eu/eli/reg/2016/679/oj}
\showURL{%
\tempurl}


\bibitem[Ficarra(2010)]%
        {DBLP:conf/hcitoch/Ficarra10}
\bibfield{author}{\bibinfo{person}{Francisco V.~Cipolla Ficarra}.}
  \bibinfo{year}{2010}\natexlab{}.
\newblock \showarticletitle{Human-Computer Interaction, Tourism and Cultural
  Heritage}. In \bibinfo{booktitle}{\emph{1st International Workshop
  Human-Computer Interaction, Tourism and Cultural Heritage}}
  \emph{(\bibinfo{series}{Lecture Notes in Computer Science},
  Vol.~\bibinfo{volume}{6529})}. \bibinfo{publisher}{Springer},
  \bibinfo{pages}{39--50}.
\newblock
\href{https://doi.org/10.1007/978-3-642-18348-5_5}{doi:\nolinkurl{10.1007/978-3-642-18348-5_5}}


\bibitem[Gavalas et~al\mbox{.}(2015)]%
        {DBLP:journals/cor/GavalasKMPV15}
\bibfield{author}{\bibinfo{person}{Damianos Gavalas},
  \bibinfo{person}{Charalampos Konstantopoulos}, \bibinfo{person}{Konstantinos
  Mastakas}, \bibinfo{person}{Grammati~E. Pantziou}, {and}
  \bibinfo{person}{Nikolaos Vathis}.} \bibinfo{year}{2015}\natexlab{}.
\newblock \showarticletitle{Heuristics for the time dependent team orienteering
  problem: Application to tourist route planning}.
\newblock \bibinfo{journal}{\emph{Computers \& Operations Research and their
  Application to Problems of World Concern}}  \bibinfo{volume}{62}
  (\bibinfo{year}{2015}), \bibinfo{pages}{36--50}.
\newblock
\href{https://doi.org/10.1016/J.COR.2015.03.016}{doi:\nolinkurl{10.1016/J.COR.2015.03.016}}


\bibitem[Gnoth(1997)]%
        {gnoth1997tourism}
\bibfield{author}{\bibinfo{person}{Juergen Gnoth}.}
  \bibinfo{year}{1997}\natexlab{}.
\newblock \showarticletitle{Tourism motivation and expectation formation}.
\newblock \bibinfo{journal}{\emph{Annals of Tourism research}}
  \bibinfo{volume}{24}, \bibinfo{number}{2} (\bibinfo{year}{1997}),
  \bibinfo{pages}{283--304}.
\newblock
\showISSN{0160-7383}
\href{https://doi.org/10.1016/s0160-7383(97)80002-3}{doi:\nolinkurl{10.1016/s0160-7383(97)80002-3}}


\bibitem[Goldenberg and Albert(2022)]%
        {Goldenberg22}
\bibfield{author}{\bibinfo{person}{Dmitri Goldenberg} {and}
  \bibinfo{person}{Javier Albert}.} \bibinfo{year}{2022}\natexlab{}.
\newblock \showarticletitle{Personalizing Benefits Allocation Without Spending
  Money: Utilizing Uplift Modeling in a Budget Constrained Setup}. In
  \bibinfo{booktitle}{\emph{16th ACM Conference on Recommender Systems}}.
  \bibinfo{publisher}{ACM}, \bibinfo{address}{New York, NY, USA},
  \bibinfo{pages}{464--465}.
\newblock
\showISBNx{9781450392785}
\href{https://doi.org/10.1145/3523227.3547381}{doi:\nolinkurl{10.1145/3523227.3547381}}


\bibitem[Goldenberg et~al\mbox{.}(2021)]%
        {DBLP:conf/rectour/GoldenbergMHKLM21}
\bibfield{author}{\bibinfo{person}{Dmitri Goldenberg}, \bibinfo{person}{Sarai
  Mizrachi}, \bibinfo{person}{Adam Horowitz}, \bibinfo{person}{Ioannis Kangas},
  \bibinfo{person}{Or Levkovich}, \bibinfo{person}{Alessandro Mozzato},
  \bibinfo{person}{Maud Schwoerer}, \bibinfo{person}{Michele Ferretti},
  \bibinfo{person}{Panagiotis Korvesis}, {and} \bibinfo{person}{Lucas
  Bernardi}.} \bibinfo{year}{2021}\natexlab{}.
\newblock \showarticletitle{I Know What You Did Next Summer: Challenges in
  Travel Destinations Recommendation}. In \bibinfo{booktitle}{\emph{RecSys
  Workshop on Recommenders in Tourism}} \emph{(\bibinfo{series}{{CEUR} Workshop
  Proceedings}, Vol.~\bibinfo{volume}{2974})}.
  \bibinfo{publisher}{CEUR-WS.org}, \bibinfo{pages}{7--22}.
\newblock


\bibitem[Gretzel(2011)]%
        {gretzel2011intelligent}
\bibfield{author}{\bibinfo{person}{Ulrike Gretzel}.}
  \bibinfo{year}{2011}\natexlab{}.
\newblock \showarticletitle{Intelligent systems in tourism: A social science
  perspective}.
\newblock \bibinfo{journal}{\emph{Annals of tourism research}}
  \bibinfo{volume}{38}, \bibinfo{number}{3} (\bibinfo{year}{2011}),
  \bibinfo{pages}{757--779}.
\newblock
\showISSN{0160-7383}
\href{https://doi.org/10.1016/j.annals.2011.04.014}{doi:\nolinkurl{10.1016/j.annals.2011.04.014}}


\bibitem[Hazwani et~al\mbox{.}(2024)]%
        {HazwaniLIREB24}
\bibfield{author}{\bibinfo{person}{Ibrahim~Al Hazwani},
  \bibinfo{person}{Tiantian Luo}, \bibinfo{person}{Oana Inel},
  \bibinfo{person}{Francesco Ricci}, \bibinfo{person}{Mennatallah El{-}Assady},
  {and} \bibinfo{person}{J{\"{u}}rgen Bernard}.}
  \bibinfo{year}{2024}\natexlab{}.
\newblock \showarticletitle{ScrollyPOI: {A} Narrative-Driven Interactive
  Recommender System for Points-of-Interest Exploration and Explainability}. In
  \bibinfo{booktitle}{\emph{32nd {ACM} Conference on User Modeling, Adaptation
  and Personalization}}. \bibinfo{publisher}{{ACM}}.
\newblock
\href{https://doi.org/10.1145/3631700.3665183}{doi:\nolinkurl{10.1145/3631700.3665183}}


\bibitem[Hernández et~al\mbox{.}(2018)]%
        {HERNANDEZ201835}
\bibfield{author}{\bibinfo{person}{Juan~M. Hernández},
  \bibinfo{person}{Andrei~P. Kirilenko}, {and} \bibinfo{person}{Svetlana
  Stepchenkova}.} \bibinfo{year}{2018}\natexlab{}.
\newblock \showarticletitle{Network approach to tourist segmentation via user
  generated content}.
\newblock \bibinfo{journal}{\emph{Annals of Tourism Research}}
  \bibinfo{volume}{73} (\bibinfo{year}{2018}), \bibinfo{pages}{35--47}.
\newblock
\showISSN{0160-7383}
\href{https://doi.org/10.1016/j.annals.2018.09.002}{doi:\nolinkurl{10.1016/j.annals.2018.09.002}}


\bibitem[Herzog et~al\mbox{.}(2019)]%
        {Herzog2019}
\bibfield{author}{\bibinfo{person}{Daniel Herzog}, \bibinfo{person}{Linus~W.
  Dietz}, {and} \bibinfo{person}{Wolfgang W\"orndl}.}
  \bibinfo{year}{2019}\natexlab{}.
\newblock \showarticletitle{Tourist Trip Recommendations -- Foundations, State
  of the Art and Challenges}.
\newblock In \bibinfo{booktitle}{\emph{Personalized Human-Computer
  Interaction}}, \bibfield{editor}{\bibinfo{person}{Miriam Augstein},
  \bibinfo{person}{Eelco Herder}, {and} \bibinfo{person}{Wolfgang W\"orndl}}
  (Eds.). \bibinfo{publisher}{de Gruyter Oldenbourg}, \bibinfo{address}{Berlin,
  Germany}, \bibinfo{pages}{159--182}.
\newblock
\showISBNx{978-3-11-055247-8}
\href{https://doi.org/10.1515/9783110552485-006}{doi:\nolinkurl{10.1515/9783110552485-006}}


\bibitem[Hofschen et~al\mbox{.}(2023)]%
        {HofschenM023}
\bibfield{author}{\bibinfo{person}{Katharina Hofschen}, \bibinfo{person}{David
  Massimo}, {and} \bibinfo{person}{Francesco Ricci}.}
  \bibinfo{year}{2023}\natexlab{}.
\newblock \showarticletitle{Expected and Experienced Utility of Points of
  Interest in Tourism Recommender Systems}. In \bibinfo{booktitle}{\emph{31st
  {ACM} Conference on User Modeling, Adaptation and Personalization}}.
  \bibinfo{publisher}{{ACM}}, \bibinfo{pages}{50--55}.
\newblock
\href{https://doi.org/10.1145/3563359.3597405}{doi:\nolinkurl{10.1145/3563359.3597405}}


\bibitem[Hutson(2018)]%
        {Hutson2018Artificial}
\bibfield{author}{\bibinfo{person}{Matthew Hutson}.}
  \bibinfo{year}{2018}\natexlab{}.
\newblock \showarticletitle{Artificial intelligence faces reproducibility
  crisis}.
\newblock \bibinfo{journal}{\emph{Science}} \bibinfo{volume}{359},
  \bibinfo{number}{6377} (\bibinfo{year}{2018}), \bibinfo{pages}{725--726}.
\newblock
\href{https://doi.org/10.1126/science.359.6377.725}{doi:\nolinkurl{10.1126/science.359.6377.725}}


\bibitem[Jameson et~al\mbox{.}(2022)]%
        {JamesonWF22}
\bibfield{author}{\bibinfo{person}{Anthony Jameson},
  \bibinfo{person}{Martijn~C. Willemsen}, {and} \bibinfo{person}{Alexander
  Felfernig}.} \bibinfo{year}{2022}\natexlab{}.
\newblock \showarticletitle{Individual and Group Decision Making and
  Recommender Systems}.
\newblock In \bibinfo{booktitle}{\emph{Recommender Systems Handbook}},
  \bibfield{editor}{\bibinfo{person}{Francesco Ricci}, \bibinfo{person}{Lior
  Rokach}, {and} \bibinfo{person}{Bracha Shapira}} (Eds.).
  \bibinfo{publisher}{Springer}, \bibinfo{pages}{789--832}.
\newblock
\href{https://doi.org/10.1007/978-1-0716-2197-4_21}{doi:\nolinkurl{10.1007/978-1-0716-2197-4_21}}


\bibitem[Jannach et~al\mbox{.}(2022)]%
        {DBLP:journals/csur/JannachMCC21}
\bibfield{author}{\bibinfo{person}{Dietmar Jannach}, \bibinfo{person}{Ahtsham
  Manzoor}, \bibinfo{person}{Wanling Cai}, {and} \bibinfo{person}{Li Chen}.}
  \bibinfo{year}{2022}\natexlab{}.
\newblock \showarticletitle{A Survey on Conversational Recommender Systems}.
\newblock \bibinfo{journal}{\emph{Comput. Surveys}} \bibinfo{volume}{54},
  \bibinfo{number}{5} (\bibinfo{year}{2022}), \bibinfo{pages}{105:1--105:36}.
\newblock
\href{https://doi.org/10.1145/3453154}{doi:\nolinkurl{10.1145/3453154}}


\bibitem[Jannach and Zanker(2020)]%
        {jannach2020interactive}
\bibfield{author}{\bibinfo{person}{Dietmar Jannach} {and}
  \bibinfo{person}{Markus Zanker}.} \bibinfo{year}{2020}\natexlab{}.
\newblock \showarticletitle{Interactive and Context-Aware Systems in Tourism}.
\newblock In \bibinfo{booktitle}{\emph{Handbook of e-Tourism}},
  \bibfield{editor}{\bibinfo{person}{Zheng Xiang}, \bibinfo{person}{Matthias
  Fuchs}, \bibinfo{person}{Ulrike Gretzel}, {and} \bibinfo{person}{Wolfram
  H{\"o}pken}} (Eds.). \bibinfo{publisher}{Springer}, \bibinfo{address}{Cham},
  \bibinfo{pages}{1--22}.
\newblock
\showISBNx{978-3-030-05324-6}
\href{https://doi.org/10.1007/978-3-030-05324-6_125-1}{doi:\nolinkurl{10.1007/978-3-030-05324-6_125-1}}


\bibitem[J{\"{a}}rv(2017)]%
        {Jarv17}
\bibfield{author}{\bibinfo{person}{Priit J{\"{a}}rv}.}
  \bibinfo{year}{2017}\natexlab{}.
\newblock \showarticletitle{Extracting Human Mobility Data from Geo-tagged
  Photos}. In \bibinfo{booktitle}{\emph{1st {ACM} {SIGSPATIAL} Workshop on
  Prediction of Human Mobility}}. \bibinfo{publisher}{{ACM}},
  \bibinfo{pages}{4:1--4:7}.
\newblock
\href{https://doi.org/10.1145/3152341.3152346}{doi:\nolinkurl{10.1145/3152341.3152346}}


\bibitem[Ji et~al\mbox{.}(2023)]%
        {DBLP:journals/tois/JiS0L23}
\bibfield{author}{\bibinfo{person}{Yitong Ji}, \bibinfo{person}{Aixin Sun},
  \bibinfo{person}{Jie Zhang}, {and} \bibinfo{person}{Chenliang Li}.}
  \bibinfo{year}{2023}\natexlab{}.
\newblock \showarticletitle{A Critical Study on Data Leakage in Recommender
  System Offline Evaluation}.
\newblock \bibinfo{journal}{\emph{{ACM} Transactions on Information Systems}}
  \bibinfo{volume}{41}, \bibinfo{number}{3} (\bibinfo{year}{2023}),
  \bibinfo{pages}{75:1--75:27}.
\newblock
\href{https://doi.org/10.1145/3569930}{doi:\nolinkurl{10.1145/3569930}}


\bibitem[Karahodža et~al\mbox{.}(2025)]%
        {Karahodza2025GroupDynamics}
\bibfield{author}{\bibinfo{person}{Esma Karahodža}, \bibinfo{person}{Amra
  Delić}, {and} \bibinfo{person}{Francesco Ricci}.}
  \bibinfo{year}{2025}\natexlab{}.
\newblock \showarticletitle{Conceptual Framework for Group Dynamics Modeling
  from Group Chat Interactions}. In \bibinfo{booktitle}{\emph{33rd ACM
  Conference on User Modeling, Adaptation and Personalization}}.
  \bibinfo{publisher}{ACM}, \bibinfo{address}{New York City, NY, USA},
  \bibinfo{pages}{5}.
\newblock
\href{https://doi.org/10.1145/3708319.3733682}{doi:\nolinkurl{10.1145/3708319.3733682}}


\bibitem[Khalil and Kobra(2021)]%
        {khalil2021monitoring}
\bibfield{author}{\bibinfo{person}{Md.~Ibrahim Khalil} {and}
  \bibinfo{person}{Mst.~Khadijatul Kobra}.} \bibinfo{year}{2021}\natexlab{}.
\newblock \showarticletitle{Monitoring overtourism: Destination Management
  Systems as a way forward}.
\newblock \bibinfo{journal}{\emph{Bangladesh Journal of Public Administration}}
  \bibinfo{volume}{29}, \bibinfo{number}{2} (\bibinfo{year}{2021}),
  \bibinfo{pages}{79--87}.
\newblock
\showISSN{1563-5023}
\href{https://doi.org/10.36609/bjpa.v29i2.211}{doi:\nolinkurl{10.36609/bjpa.v29i2.211}}


\bibitem[Kim et~al\mbox{.}(2021)]%
        {Kim2021Event}
\bibfield{author}{\bibinfo{person}{Minkyoung Kim}, \bibinfo{person}{Lexing
  Xie}, {and} \bibinfo{person}{Peter Christen}.}
  \bibinfo{year}{2021}\natexlab{}.
\newblock \showarticletitle{Event Diffusion Patterns in Social Media}.
\newblock \bibinfo{journal}{\emph{International AAAI Conference on Web and
  Social Media}} \bibinfo{volume}{6}, \bibinfo{number}{1} (\bibinfo{date}{Aug.}
  \bibinfo{year}{2021}), \bibinfo{pages}{178--185}.
\newblock
\showISSN{2162-3449}
\href{https://doi.org/10.1609/icwsm.v6i1.14248}{doi:\nolinkurl{10.1609/icwsm.v6i1.14248}}


\bibitem[Leal et~al\mbox{.}(2018)]%
        {leal2018context}
\bibfield{author}{\bibinfo{person}{Fátima Leal}, \bibinfo{person}{Benedita
  Malheiro}, {and} \bibinfo{person}{Juan~C. Burguillo}.}
  \bibinfo{year}{2018}\natexlab{}.
\newblock \showarticletitle{Context-aware tourism technologies}.
\newblock \bibinfo{journal}{\emph{The Knowledge Engineering Review}}
  \bibinfo{volume}{33} (\bibinfo{year}{2018}).
\newblock
\showISSN{1469-8005}
\href{https://doi.org/10.1017/s0269888918000152}{doi:\nolinkurl{10.1017/s0269888918000152}}


\bibitem[Leal et~al\mbox{.}(2025)]%
        {DBLP:journals/es/LealVMB25}
\bibfield{author}{\bibinfo{person}{F{\'{a}}tima Leal}, \bibinfo{person}{Bruno
  Veloso}, \bibinfo{person}{Benedita Malheiro}, {and} \bibinfo{person}{Juan~C.
  Burguillo}.} \bibinfo{year}{2025}\natexlab{}.
\newblock \showarticletitle{Towards adaptive and transparent tourism
  recommendations: {A} survey}.
\newblock \bibinfo{journal}{\emph{Expert Systems: The Journal of Knowledge
  Engineering}} \bibinfo{volume}{42}, \bibinfo{number}{1}
  (\bibinfo{year}{2025}).
\newblock
\href{https://doi.org/10.1111/EXSY.13400}{doi:\nolinkurl{10.1111/EXSY.13400}}


\bibitem[Lestari et~al\mbox{.}(2022)]%
        {pgr}
\bibfield{author}{\bibinfo{person}{Forina Lestari},
  \bibinfo{person}{Melasutra~Md Dali}, {and} \bibinfo{person}{Norbani Che-Ha}.}
  \bibinfo{year}{2022}\natexlab{}.
\newblock \showarticletitle{The Importance of A Multistakeholder Perspective in
  Mapping Stakeholders' Roles Toward City Branding Implementation}.
\newblock \bibinfo{journal}{\emph{Policy \& Governance Review}}
  \bibinfo{volume}{6}, \bibinfo{number}{3} (\bibinfo{year}{2022}),
  \bibinfo{pages}{264--281}.
\newblock
\showISSN{2580-4820}
\href{https://doi.org/10.30589/pgr.v6i3.601}{doi:\nolinkurl{10.30589/pgr.v6i3.601}}


\bibitem[Li et~al\mbox{.}(2021)]%
        {DBLP:conf/dasfaa/LiHZZLS21}
\bibfield{author}{\bibinfo{person}{Changheng Li}, \bibinfo{person}{Yongjing
  Hao}, \bibinfo{person}{Pengpeng Zhao}, \bibinfo{person}{Fuzhen Zhuang},
  \bibinfo{person}{Yanchi Liu}, {and} \bibinfo{person}{Victor~S. Sheng}.}
  \bibinfo{year}{2021}\natexlab{}.
\newblock \showarticletitle{Tell Me Where to Go Next: Improving {POI}
  Recommendation via Conversation}. In \bibinfo{booktitle}{\emph{26th
  International Conference on Database Systems for Advanced Applications}}
  \emph{(\bibinfo{series}{Lecture Notes in Computer Science},
  Vol.~\bibinfo{volume}{12683})}. \bibinfo{publisher}{Springer},
  \bibinfo{pages}{211--227}.
\newblock
\href{https://doi.org/10.1007/978-3-030-73200-4_14}{doi:\nolinkurl{10.1007/978-3-030-73200-4_14}}


\bibitem[Li et~al\mbox{.}(2024)]%
        {DBLP:conf/sigir/LiR0A0S24}
\bibfield{author}{\bibinfo{person}{Peibo Li}, \bibinfo{person}{Maarten de
  Rijke}, \bibinfo{person}{Hao Xue}, \bibinfo{person}{Shuang Ao},
  \bibinfo{person}{Yang Song}, {and} \bibinfo{person}{Flora~D. Salim}.}
  \bibinfo{year}{2024}\natexlab{}.
\newblock \showarticletitle{Large Language Models for Next Point-of-Interest
  Recommendation}. In \bibinfo{booktitle}{\emph{47th International ACM SIGIR
  Conference on Research and Development in Information Retrieval}}
  \emph{(\bibinfo{series}{SIGIR'24})}. \bibinfo{publisher}{{ACM}},
  \bibinfo{pages}{1463--1472}.
\newblock
\href{https://doi.org/10.1145/3626772.3657840}{doi:\nolinkurl{10.1145/3626772.3657840}}


\bibitem[Lim et~al\mbox{.}(2016)]%
        {DBLP:conf/aips/LimCLK16}
\bibfield{author}{\bibinfo{person}{Kwan~Hui Lim}, \bibinfo{person}{Jeffrey
  Chan}, \bibinfo{person}{Christopher Leckie}, {and} \bibinfo{person}{Shanika
  Karunasekera}.} \bibinfo{year}{2016}\natexlab{}.
\newblock \showarticletitle{Towards Next Generation Touring: Personalized Group
  Tours}. In \bibinfo{booktitle}{\emph{26th International Conference on
  Automated Planning and Scheduling}}, Vol.~\bibinfo{volume}{26}.
  \bibinfo{publisher}{{AAAI}}, \bibinfo{pages}{412--420}.
\newblock
\showISSN{2334-0835}
\href{https://doi.org/10.1609/icaps.v26i1.13775}{doi:\nolinkurl{10.1609/icaps.v26i1.13775}}


\bibitem[Lim et~al\mbox{.}(2018)]%
        {DBLP:journals/kais/LimCLK18}
\bibfield{author}{\bibinfo{person}{Kwan~Hui Lim}, \bibinfo{person}{Jeffrey
  Chan}, \bibinfo{person}{Christopher Leckie}, {and} \bibinfo{person}{Shanika
  Karunasekera}.} \bibinfo{year}{2018}\natexlab{}.
\newblock \showarticletitle{Personalized trip recommendation for tourists based
  on user interests, points of interest visit durations and visit recency}.
\newblock \bibinfo{journal}{\emph{Knowledge and Information Systems}}
  \bibinfo{volume}{54}, \bibinfo{number}{2} (\bibinfo{year}{2018}),
  \bibinfo{pages}{375--406}.
\newblock
\href{https://doi.org/10.1007/S10115-017-1056-Y}{doi:\nolinkurl{10.1007/S10115-017-1056-Y}}


\bibitem[Linxen et~al\mbox{.}(2021)]%
        {Linxen2021Weird}
\bibfield{author}{\bibinfo{person}{Sebastian Linxen},
  \bibinfo{person}{Christian Sturm}, \bibinfo{person}{Florian Br\"{u}hlmann},
  \bibinfo{person}{Vincent Cassau}, \bibinfo{person}{Klaus Opwis}, {and}
  \bibinfo{person}{Katharina Reinecke}.} \bibinfo{year}{2021}\natexlab{}.
\newblock \showarticletitle{How WEIRD is CHI?}. In
  \bibinfo{booktitle}{\emph{2021 CHI Conference on Human Factors in Computing
  Systems}} \emph{(\bibinfo{series}{CHI '21})}. \bibinfo{publisher}{ACM},
  \bibinfo{address}{New York, NY, USA}, Article \bibinfo{articleno}{143},
  \bibinfo{numpages}{14}~pages.
\newblock
\showISBNx{9781450380966}
\href{https://doi.org/10.1145/3411764.3445488}{doi:\nolinkurl{10.1145/3411764.3445488}}


\bibitem[Liu et~al\mbox{.}(2022)]%
        {DBLP:journals/ijon/LiuYXYHW22}
\bibfield{author}{\bibinfo{person}{Xin Liu}, \bibinfo{person}{Yongjian Yang},
  \bibinfo{person}{Yuanbo Xu}, \bibinfo{person}{Funing Yang},
  \bibinfo{person}{Qiuyang Huang}, {and} \bibinfo{person}{Hong Wang}.}
  \bibinfo{year}{2022}\natexlab{}.
\newblock \showarticletitle{Real-time {POI} recommendation via modeling long-
  and short-term user preferences}.
\newblock \bibinfo{journal}{\emph{Neurocomputing}}  \bibinfo{volume}{467}
  (\bibinfo{year}{2022}), \bibinfo{pages}{454--464}.
\newblock
\href{https://doi.org/10.1016/J.NEUCOM.2021.09.056}{doi:\nolinkurl{10.1016/J.NEUCOM.2021.09.056}}


\bibitem[Madeira et~al\mbox{.}(2024)]%
        {Neves24}
\bibfield{author}{\bibinfo{person}{Rui~Neves Madeira}, \bibinfo{person}{Samuel
  Robalo}, {and} \bibinfo{person}{Andr\'{e} Cordeiro}.}
  \bibinfo{year}{2024}\natexlab{}.
\newblock \showarticletitle{SHIFT towards Sustainable Tourism: Designing a
  Ubiquitous Platform for Co-Creating Experiences with Tourists and Residents}.
  In \bibinfo{booktitle}{\emph{International Conference on Mobile and
  Ubiquitous Multimedia}}. \bibinfo{publisher}{ACM}, \bibinfo{address}{New
  York, NY, USA}, \bibinfo{pages}{467--470}.
\newblock
\showISBNx{9798400712838}
\href{https://doi.org/10.1145/3701571.3703387}{doi:\nolinkurl{10.1145/3701571.3703387}}


\bibitem[Mao et~al\mbox{.}(2024)]%
        {muca}
\bibfield{author}{\bibinfo{person}{Manqing Mao}, \bibinfo{person}{Paishun
  Ting}, \bibinfo{person}{Yijian Xiang}, \bibinfo{person}{Mingyang Xu},
  \bibinfo{person}{Julia Chen}, {and} \bibinfo{person}{Jianzhe Lin}.}
  \bibinfo{year}{2024}\natexlab{}.
\newblock \showarticletitle{Multi-User Chat Assistant ({MUCA}): a Framework
  Using {LLMs} to Facilitate Group Conversations}.
\newblock \bibinfo{journal}{\emph{arXiv}} (\bibinfo{year}{2024}).
\newblock
\href{https://doi.org/10.48550/arXiv.2401.04883}{doi:\nolinkurl{10.48550/arXiv.2401.04883}}


\bibitem[Massimo and Ricci(2019)]%
        {DBLP:conf/rectour/Massimo019}
\bibfield{author}{\bibinfo{person}{David Massimo} {and}
  \bibinfo{person}{Francesco Ricci}.} \bibinfo{year}{2019}\natexlab{}.
\newblock \showarticletitle{Users' Evaluation of Next-POI Recommendations}. In
  \bibinfo{booktitle}{\emph{RecSys Recommenders in Tourism Workshop}}
  \emph{(\bibinfo{series}{{CEUR} Workshop Proceedings},
  Vol.~\bibinfo{volume}{2435})}. \bibinfo{publisher}{CEUR-WS.org},
  \bibinfo{pages}{1--8}.
\newblock


\bibitem[Massimo and Ricci(2021)]%
        {MassimoR21}
\bibfield{author}{\bibinfo{person}{David Massimo} {and}
  \bibinfo{person}{Francesco Ricci}.} \bibinfo{year}{2021}\natexlab{}.
\newblock \showarticletitle{Popularity, novelty and relevance in point of
  interest recommendation: an experimental analysis}.
\newblock \bibinfo{journal}{\emph{Information Technology \& Tourism}}
  \bibinfo{volume}{23}, \bibinfo{number}{4} (\bibinfo{year}{2021}),
  \bibinfo{pages}{473--508}.
\newblock
\href{https://doi.org/10.1007/S40558-021-00214-5}{doi:\nolinkurl{10.1007/S40558-021-00214-5}}


\bibitem[Massimo and Ricci(2023)]%
        {Massimo023}
\bibfield{author}{\bibinfo{person}{David Massimo} {and}
  \bibinfo{person}{Francesco Ricci}.} \bibinfo{year}{2023}\natexlab{}.
\newblock \showarticletitle{Combining Reinforcement Learning and Spatial
  Proximity Exploration for New User and New {POI} Recommendations}. In
  \bibinfo{booktitle}{\emph{31st {ACM} Conference on User Modeling, Adaptation
  and Personalization}}. \bibinfo{publisher}{{ACM}}, \bibinfo{pages}{164--174}.
\newblock
\href{https://doi.org/10.1145/3565472.3592966}{doi:\nolinkurl{10.1145/3565472.3592966}}


\bibitem[Mauro et~al\mbox{.}(2024)]%
        {Mauro24}
\bibfield{author}{\bibinfo{person}{Noemi Mauro}, \bibinfo{person}{Livio
  Scarpinati}, \bibinfo{person}{Fabio Ferrero}, \bibinfo{person}{Angelo
  Geninatti~Cossatin}, {and} \bibinfo{person}{Claudio Mattutino}.}
  \bibinfo{year}{2024}\natexlab{}.
\newblock \showarticletitle{Point-of-Interest Recommender Systems: Nudging
  towards Sustainable Tourism}. In \bibinfo{booktitle}{\emph{32nd ACM
  Conference on User Modeling, Adaptation and Personalization}}.
  \bibinfo{publisher}{ACM}, \bibinfo{address}{New York, NY, USA},
  \bibinfo{pages}{491--495}.
\newblock
\showISBNx{9798400704666}
\href{https://doi.org/10.1145/3631700.3664904}{doi:\nolinkurl{10.1145/3631700.3664904}}


\bibitem[Meng et~al\mbox{.}(2020)]%
        {DBLP:conf/recsys/MengMMO20}
\bibfield{author}{\bibinfo{person}{Zaiqiao Meng}, \bibinfo{person}{Richard
  McCreadie}, \bibinfo{person}{Craig Macdonald}, {and} \bibinfo{person}{Iadh
  Ounis}.} \bibinfo{year}{2020}\natexlab{}.
\newblock \showarticletitle{Exploring Data Splitting Strategies for the
  Evaluation of Recommendation Models}. In \bibinfo{booktitle}{\emph{14th {ACM}
  Conference on Recommender Systems}}. \bibinfo{publisher}{{ACM}},
  \bibinfo{pages}{681--686}.
\newblock
\href{https://doi.org/10.1145/3383313.3418479}{doi:\nolinkurl{10.1145/3383313.3418479}}


\bibitem[Merinov et~al\mbox{.}(2023)]%
        {MerinovM023}
\bibfield{author}{\bibinfo{person}{Pavel Merinov}, \bibinfo{person}{David
  Massimo}, {and} \bibinfo{person}{Francesco Ricci}.}
  \bibinfo{year}{2023}\natexlab{}.
\newblock \showarticletitle{Behaviour-aware Tourist Profiles Data Generation}.
  In \bibinfo{booktitle}{\emph{13th Italian Information Retrieval Workshop}}
  \emph{(\bibinfo{series}{{CEUR} Workshop Proceedings},
  Vol.~\bibinfo{volume}{3448})}. \bibinfo{publisher}{CEUR-WS.org},
  \bibinfo{pages}{3--8}.
\newblock


\bibitem[Merinov and Ricci(2024)]%
        {Merinov024}
\bibfield{author}{\bibinfo{person}{Pavel Merinov} {and}
  \bibinfo{person}{Francesco Ricci}.} \bibinfo{year}{2024}\natexlab{}.
\newblock \showarticletitle{Positive-Sum Impact of Multistakeholder Recommender
  Systems for Urban Tourism Promotion and User Utility}. In
  \bibinfo{booktitle}{\emph{18th {ACM} Conference on Recommender Systems}}.
  \bibinfo{publisher}{{ACM}}, \bibinfo{pages}{939--944}.
\newblock
\href{https://doi.org/10.1145/3640457.3688173}{doi:\nolinkurl{10.1145/3640457.3688173}}


\bibitem[Migliorini et~al\mbox{.}(2021)]%
        {DBLP:journals/tetc/MiglioriniCB21}
\bibfield{author}{\bibinfo{person}{Sara Migliorini}, \bibinfo{person}{Damiano
  Carra}, {and} \bibinfo{person}{Alberto Belussi}.}
  \bibinfo{year}{2021}\natexlab{}.
\newblock \showarticletitle{Distributing Tourists among POIs with an Adaptive
  Trip Recommendation System}.
\newblock \bibinfo{journal}{\emph{{IEEE} Transactions on Emerging Topics in
  Computing}} \bibinfo{volume}{9}, \bibinfo{number}{4} (\bibinfo{year}{2021}),
  \bibinfo{pages}{1765--1779}.
\newblock
\href{https://doi.org/10.1109/TETC.2019.2920484}{doi:\nolinkurl{10.1109/TETC.2019.2920484}}


\bibitem[Nguyen and Ricci(2018)]%
        {Nguyen018}
\bibfield{author}{\bibinfo{person}{Thuy~Ngoc Nguyen} {and}
  \bibinfo{person}{Francesco Ricci}.} \bibinfo{year}{2018}\natexlab{}.
\newblock \showarticletitle{A chat-based group recommender system for tourism}.
\newblock \bibinfo{journal}{\emph{Information Technology \& Tourism}}
  \bibinfo{volume}{18}, \bibinfo{number}{1-4} (\bibinfo{year}{2018}),
  \bibinfo{pages}{5--28}.
\newblock
\href{https://doi.org/10.1007/S40558-017-0099-Y}{doi:\nolinkurl{10.1007/S40558-017-0099-Y}}


\bibitem[Nguyen et~al\mbox{.}(2019)]%
        {NguyenRDB19}
\bibfield{author}{\bibinfo{person}{Thuy~Ngoc Nguyen},
  \bibinfo{person}{Francesco Ricci}, \bibinfo{person}{Amra Delic}, {and}
  \bibinfo{person}{Derek~G. Bridge}.} \bibinfo{year}{2019}\natexlab{}.
\newblock \showarticletitle{Conflict resolution in group decision making:
  insights from a simulation study}.
\newblock \bibinfo{journal}{\emph{User Modeling and User-Adapted Interaction}}
  \bibinfo{volume}{29}, \bibinfo{number}{5} (\bibinfo{year}{2019}),
  \bibinfo{pages}{895--941}.
\newblock
\href{https://doi.org/10.1007/S11257-019-09240-9}{doi:\nolinkurl{10.1007/S11257-019-09240-9}}


\bibitem[Noulas et~al\mbox{.}(2011)]%
        {DBLP:conf/icwsm/NoulasSMP11}
\bibfield{author}{\bibinfo{person}{Anastasios Noulas},
  \bibinfo{person}{Salvatore Scellato}, \bibinfo{person}{Cecilia Mascolo},
  {and} \bibinfo{person}{Massimiliano Pontil}.}
  \bibinfo{year}{2011}\natexlab{}.
\newblock \showarticletitle{An Empirical Study of Geographic User Activity
  Patterns in Foursquare}. In \bibinfo{booktitle}{\emph{5th International
  Conference on Weblogs and Social Media}}, Vol.~\bibinfo{volume}{5}.
  \bibinfo{publisher}{{AAAI}}, \bibinfo{pages}{570--573}.
\newblock
\showISSN{2162-3449}
\href{https://doi.org/10.1609/icwsm.v5i1.14175}{doi:\nolinkurl{10.1609/icwsm.v5i1.14175}}


\bibitem[Orabi et~al\mbox{.}(2020)]%
        {Orabi2020Detection}
\bibfield{author}{\bibinfo{person}{Mariam Orabi}, \bibinfo{person}{Djedjiga
  Mouheb}, \bibinfo{person}{Zaher Al~Aghbari}, {and} \bibinfo{person}{Ibrahim
  Kamel}.} \bibinfo{year}{2020}\natexlab{}.
\newblock \showarticletitle{Detection of bots in social media: a systematic
  review}.
\newblock \bibinfo{journal}{\emph{Information Processing \& Management}}
  \bibinfo{volume}{57}, \bibinfo{number}{4} (\bibinfo{date}{July}
  \bibinfo{year}{2020}), \bibinfo{pages}{102250}.
\newblock
\showISSN{0306-4573}
\href{https://doi.org/10.1016/j.ipm.2020.102250}{doi:\nolinkurl{10.1016/j.ipm.2020.102250}}


\bibitem[Otaki and Baba(2025)]%
        {DBLP:journals/eswa/OtakiB25}
\bibfield{author}{\bibinfo{person}{Keisuke Otaki} {and} \bibinfo{person}{Yukino
  Baba}.} \bibinfo{year}{2025}\natexlab{}.
\newblock \showarticletitle{Travel itinerary recommendation using
  interaction-based augmented data}.
\newblock \bibinfo{journal}{\emph{Expert Systems with Applications}}
  \bibinfo{volume}{269} (\bibinfo{year}{2025}), \bibinfo{pages}{126294}.
\newblock
\href{https://doi.org/10.1016/J.ESWA.2024.126294}{doi:\nolinkurl{10.1016/J.ESWA.2024.126294}}


\bibitem[Ozcelik et~al\mbox{.}(2024)]%
        {Suayb23}
\bibfield{author}{\bibinfo{person}{Suayb~Talha Ozcelik},
  \bibinfo{person}{Meltem~Turhan Yondem}, \bibinfo{person}{Ines Caetano},
  \bibinfo{person}{Jose Figueiredo}, \bibinfo{person}{Patricia Alves},
  \bibinfo{person}{Goreti Marreiros}, \bibinfo{person}{Huseyin Bahtiyar},
  \bibinfo{person}{Eda Yuksel}, \bibinfo{person}{Fernando Perales}, {and}
  \bibinfo{person}{George Suciu}.} \bibinfo{year}{2024}\natexlab{}.
\newblock \showarticletitle{Transforming Tourism Experience: AI-Based Smart
  Travel Platform}. In \bibinfo{booktitle}{\emph{4th European Symposium on
  Software Engineering}}. \bibinfo{publisher}{ACM}, \bibinfo{address}{New York,
  NY, USA}, \bibinfo{pages}{37--45}.
\newblock
\showISBNx{9798400708817}
\href{https://doi.org/10.1145/3651640.3651645}{doi:\nolinkurl{10.1145/3651640.3651645}}


\bibitem[Park et~al\mbox{.}(2019)]%
        {park2019travel}
\bibfield{author}{\bibinfo{person}{Sangwon Park}, \bibinfo{person}{Yang Yang},
  {and} \bibinfo{person}{Mingshu Wang}.} \bibinfo{year}{2019}\natexlab{}.
\newblock \showarticletitle{Travel distance and hotel service satisfaction: An
  inverted U-shaped relationship}.
\newblock \bibinfo{journal}{\emph{International Journal of Hospitality
  Management}}  \bibinfo{volume}{76} (\bibinfo{year}{2019}),
  \bibinfo{pages}{261--270}.
\newblock
\showISSN{0278-4319}
\href{https://doi.org/10.1016/j.ijhm.2018.05.015}{doi:\nolinkurl{10.1016/j.ijhm.2018.05.015}}


\bibitem[Paulino et~al\mbox{.}(2021)]%
        {paulino2021identifying}
\bibfield{author}{\bibinfo{person}{Isabel Paulino}, \bibinfo{person}{Sergi
  Lozano}, {and} \bibinfo{person}{Lluís Prats}.}
  \bibinfo{year}{2021}\natexlab{}.
\newblock \showarticletitle{Identifying tourism destinations from tourists’
  travel patterns}.
\newblock \bibinfo{journal}{\emph{Journal of Destination Marketing \&
  Management}}  \bibinfo{volume}{19} (\bibinfo{year}{2021}),
  \bibinfo{pages}{100508}.
\newblock
\showISSN{2212-571X}
\href{https://doi.org/10.1016/j.jdmm.2020.100508}{doi:\nolinkurl{10.1016/j.jdmm.2020.100508}}


\bibitem[Pearce and Lee(2005)]%
        {pearce2005developing}
\bibfield{author}{\bibinfo{person}{Philip~L. Pearce} {and}
  \bibinfo{person}{Uk-Il Lee}.} \bibinfo{year}{2005}\natexlab{}.
\newblock \showarticletitle{Developing the Travel Career Approach to Tourist
  Motivation}.
\newblock \bibinfo{journal}{\emph{Journal of Travel Research}}
  \bibinfo{volume}{43}, \bibinfo{number}{3} (\bibinfo{year}{2005}),
  \bibinfo{pages}{226--237}.
\newblock
\href{https://doi.org/10.1177/0047287504272020}{doi:\nolinkurl{10.1177/0047287504272020}}


\bibitem[Pechlaner(2000)]%
        {pechlaner2000cultural}
\bibfield{author}{\bibinfo{person}{Harald Pechlaner}.}
  \bibinfo{year}{2000}\natexlab{}.
\newblock \showarticletitle{Cultural heritage and destination management in the
  Mediterranean}.
\newblock \bibinfo{journal}{\emph{Thunderbird International Business Review}}
  \bibinfo{volume}{42}, \bibinfo{number}{4} (\bibinfo{year}{2000}),
  \bibinfo{pages}{409--426}.
\newblock
\href{https://doi.org/10.1002/1520-6874(200007/08)42:4<409::aid-tie4>3.0.co;2-f}{doi:\nolinkurl{10.1002/1520-6874(200007/08)42:4<409::aid-tie4>3.0.co;2-f}}


\bibitem[Pechlaner et~al\mbox{.}(2019)]%
        {pechlaner2019overtourism}
\bibfield{author}{\bibinfo{person}{Harald Pechlaner}, \bibinfo{person}{Elisa
  Innerhofer}, {and} \bibinfo{person}{Greta Erschbamer}.}
  \bibinfo{year}{2019}\natexlab{}.
\newblock \bibinfo{booktitle}{\emph{Overtourism: Tourism management and
  solutions}}.
\newblock \bibinfo{publisher}{Routledge}, \bibinfo{address}{London}.
\newblock
\showISBNx{9780367187439}


\bibitem[Perry(2023)]%
        {perry2023role}
\bibfield{author}{\bibinfo{person}{Morgan Perry}.}
  \bibinfo{year}{2023}\natexlab{}.
\newblock \showarticletitle{Role of Cultural Heritage Preservation in
  Destination Branding and Tourist Experience Enhancement}.
\newblock \bibinfo{journal}{\emph{Hospitality and Tourism Journal}}
  \bibinfo{volume}{1}, \bibinfo{number}{1} (\bibinfo{year}{2023}),
  \bibinfo{pages}{36--47}.
\newblock


\bibitem[Piliponyte et~al\mbox{.}(2023)]%
        {DBLP:conf/um/PiliponyteM023}
\bibfield{author}{\bibinfo{person}{Greta Piliponyte}, \bibinfo{person}{David
  Massimo}, {and} \bibinfo{person}{Francesco Ricci}.}
  \bibinfo{year}{2023}\natexlab{}.
\newblock \showarticletitle{The Impact of Personalised Advertisement Campaigns
  on Tourist Choices in South Tyrol: {A} Sustainable Tourism Perspective}. In
  \bibinfo{booktitle}{\emph{31st {ACM} Conference on User Modeling, Adaptation
  and Personalization}}. \bibinfo{publisher}{{ACM}}, \bibinfo{pages}{100--103}.
\newblock
\href{https://doi.org/10.1145/3563359.3597445}{doi:\nolinkurl{10.1145/3563359.3597445}}


\bibitem[Piliponyte et~al\mbox{.}(2024)]%
        {PiliponyteMR24}
\bibfield{author}{\bibinfo{person}{Greta Piliponyte}, \bibinfo{person}{David
  Massimo}, {and} \bibinfo{person}{Francesco Ricci}.}
  \bibinfo{year}{2024}\natexlab{}.
\newblock \showarticletitle{Simulation of recommender systems driven tourism
  promotion campaigns}.
\newblock \bibinfo{journal}{\emph{Information Technology \& Tourism}}
  \bibinfo{volume}{26}, \bibinfo{number}{3} (\bibinfo{year}{2024}),
  \bibinfo{pages}{407--448}.
\newblock
\href{https://doi.org/10.1007/S40558-024-00283-2}{doi:\nolinkurl{10.1007/S40558-024-00283-2}}


\bibitem[Plog(1974)]%
        {doi:10.1177/001088047401400409}
\bibfield{author}{\bibinfo{person}{Stanley~C. Plog}.}
  \bibinfo{year}{1974}\natexlab{}.
\newblock \showarticletitle{Why Destination Areas Rise and Fall in Popularity}.
\newblock \bibinfo{journal}{\emph{Cornell Hotel and Restaurant Administration
  Quarterly}} \bibinfo{volume}{14}, \bibinfo{number}{4} (\bibinfo{year}{1974}),
  \bibinfo{pages}{55--58}.
\newblock
\href{https://doi.org/10.1177/001088047401400409}{doi:\nolinkurl{10.1177/001088047401400409}}


\bibitem[Ricci and Delić(2025)]%
        {ricci2025CAJO}
\bibfield{author}{\bibinfo{person}{Francesco Ricci} {and} \bibinfo{person}{Amra
  Delić}.} \bibinfo{year}{2025}\natexlab{}.
\newblock \bibinfo{title}{Widening the Role of Group Recommender Systems with
  CAJO}.
\newblock
\href{https://doi.org/10.48550/arxiv.2504.05934}{doi:\nolinkurl{10.48550/arxiv.2504.05934}}
\showeprint[arxiv]{2504.05934}


\bibitem[Ricci et~al\mbox{.}(2021)]%
        {RicciMA21}
\bibfield{author}{\bibinfo{person}{Francesco Ricci}, \bibinfo{person}{David
  Massimo}, {and} \bibinfo{person}{Antonella~De Angeli}.}
  \bibinfo{year}{2021}\natexlab{}.
\newblock \showarticletitle{Challenges for Recommender Systems Evaluation}. In
  \bibinfo{booktitle}{\emph{14th Biannual Conference of the Italian {SIGCHI}
  Chapter}}. \bibinfo{publisher}{{ACM}}, \bibinfo{pages}{25:1--25:5}.
\newblock
\href{https://doi.org/10.1145/3464385.3464733}{doi:\nolinkurl{10.1145/3464385.3464733}}


\bibitem[Robina‐Ramírez et~al\mbox{.}(2024)]%
        {ROBINARAMIREZ202412639}
\bibfield{author}{\bibinfo{person}{Rafael Robina‐Ramírez},
  \bibinfo{person}{Jesús Torrecilla‐Pinero}, \bibinfo{person}{Ana
  Leal‐Solís}, {and} \bibinfo{person}{Juan~Antonio Pavón‐Pérez}.}
  \bibinfo{year}{2024}\natexlab{}.
\newblock \showarticletitle{Tourism as a driver of economic and social
  development in underdeveloped regions}.
\newblock \bibinfo{journal}{\emph{Regional Science Policy \& Practice}}
  \bibinfo{volume}{16}, \bibinfo{number}{1} (\bibinfo{year}{2024}),
  \bibinfo{pages}{12639}.
\newblock
\showISSN{1757-7802}
\href{https://doi.org/10.1111/rsp3.12639}{doi:\nolinkurl{10.1111/rsp3.12639}}


\bibitem[Rossi et~al\mbox{.}(2015)]%
        {DBLP:conf/icwsm/0004WSM15}
\bibfield{author}{\bibinfo{person}{Luca Rossi}, \bibinfo{person}{Matthew
  Williams}, \bibinfo{person}{Christoph Stich}, {and} \bibinfo{person}{Mirco
  Musolesi}.} \bibinfo{year}{2015}\natexlab{}.
\newblock \showarticletitle{Privacy and the City: User Identification and
  Location Semantics in Location-Based Social Networks}. In
  \bibinfo{booktitle}{\emph{9th International Conference on Web and Social
  Media}}, Vol.~\bibinfo{volume}{9}. \bibinfo{publisher}{{AAAI}},
  \bibinfo{pages}{387--396}.
\newblock
\showISSN{2162-3449}
\href{https://doi.org/10.1609/icwsm.v9i1.14595}{doi:\nolinkurl{10.1609/icwsm.v9i1.14595}}


\bibitem[S\'{a}nchez(2019)]%
        {Sanchez19}
\bibfield{author}{\bibinfo{person}{Pablo S\'{a}nchez}.}
  \bibinfo{year}{2019}\natexlab{}.
\newblock \showarticletitle{Exploiting contextual information for recommender
  systems oriented to tourism}. In \bibinfo{booktitle}{\emph{13th ACM
  Conference on Recommender Systems}}. \bibinfo{publisher}{ACM},
  \bibinfo{address}{New York, NY, USA}, \bibinfo{pages}{601--605}.
\newblock
\showISBNx{9781450362436}
\href{https://doi.org/10.1145/3298689.3347062}{doi:\nolinkurl{10.1145/3298689.3347062}}


\bibitem[S{\'{a}}nchez and Bellog{\'{\i}}n(2020)]%
        {DBLP:journals/umuai/SanchezB20}
\bibfield{author}{\bibinfo{person}{Pablo S{\'{a}}nchez} {and}
  \bibinfo{person}{Alejandro Bellog{\'{\i}}n}.}
  \bibinfo{year}{2020}\natexlab{}.
\newblock \showarticletitle{Applying reranking strategies to route
  recommendation using sequence-aware evaluation}.
\newblock \bibinfo{journal}{\emph{User Modeling and User-Adapted Interaction}}
  \bibinfo{volume}{30}, \bibinfo{number}{4} (\bibinfo{year}{2020}),
  \bibinfo{pages}{659--725}.
\newblock
\href{https://doi.org/10.1007/S11257-020-09258-4}{doi:\nolinkurl{10.1007/S11257-020-09258-4}}


\bibitem[S{\'{a}}nchez and Bellog{\'{\i}}n(2021)]%
        {DBLP:journals/ipm/SanchezB21}
\bibfield{author}{\bibinfo{person}{Pablo S{\'{a}}nchez} {and}
  \bibinfo{person}{Alejandro Bellog{\'{\i}}n}.}
  \bibinfo{year}{2021}\natexlab{}.
\newblock \showarticletitle{On the effects of aggregation strategies for
  different groups of users in venue recommendation}.
\newblock \bibinfo{journal}{\emph{Information Processing and Management}}
  \bibinfo{volume}{58}, \bibinfo{number}{5} (\bibinfo{year}{2021}),
  \bibinfo{pages}{102609}.
\newblock
\href{https://doi.org/10.1016/J.IPM.2021.102609}{doi:\nolinkurl{10.1016/J.IPM.2021.102609}}


\bibitem[S{\'{a}}nchez and Bellog{\'{\i}}n(2022)]%
        {DBLP:journals/csur/SanchezB22}
\bibfield{author}{\bibinfo{person}{Pablo S{\'{a}}nchez} {and}
  \bibinfo{person}{Alejandro Bellog{\'{\i}}n}.}
  \bibinfo{year}{2022}\natexlab{}.
\newblock \showarticletitle{Point-of-Interest Recommender Systems Based on
  Location-Based Social Networks: {A} Survey from an Experimental Perspective}.
\newblock \bibinfo{journal}{\emph{Comput. Surveys}} \bibinfo{volume}{54},
  \bibinfo{number}{11s} (\bibinfo{year}{2022}), \bibinfo{pages}{223:1--223:37}.
\newblock
\href{https://doi.org/10.1145/3510409}{doi:\nolinkurl{10.1145/3510409}}


\bibitem[S{\'{a}}nchez et~al\mbox{.}(2023)]%
        {DBLP:journals/datamine/SanchezBB23}
\bibfield{author}{\bibinfo{person}{Pablo S{\'{a}}nchez},
  \bibinfo{person}{Alejandro Bellog{\'{\i}}n}, {and} \bibinfo{person}{Ludovico
  Boratto}.} \bibinfo{year}{2023}\natexlab{}.
\newblock \showarticletitle{Bias characterization, assessment, and mitigation
  in location-based recommender systems}.
\newblock \bibinfo{journal}{\emph{Data Mining and Knowledge Discovery}}
  \bibinfo{volume}{37}, \bibinfo{number}{5} (\bibinfo{year}{2023}),
  \bibinfo{pages}{1885--1929}.
\newblock
\href{https://doi.org/10.1007/S10618-022-00913-5}{doi:\nolinkurl{10.1007/S10618-022-00913-5}}


\bibitem[S{\'{a}}nchez and Dietz(2022)]%
        {DBLP:conf/um/SanchezD22}
\bibfield{author}{\bibinfo{person}{Pablo S{\'{a}}nchez} {and}
  \bibinfo{person}{Linus~W. Dietz}.} \bibinfo{year}{2022}\natexlab{}.
\newblock \showarticletitle{Travelers vs. Locals: The Effect of Cluster
  Analysis in Point-of-Interest Recommendation}. In
  \bibinfo{booktitle}{\emph{30th {ACM} Conference on User Modeling, Adaptation
  and Personalization}}. \bibinfo{publisher}{{ACM}}, \bibinfo{pages}{132--142}.
\newblock
\href{https://doi.org/10.1145/3503252.3531320}{doi:\nolinkurl{10.1145/3503252.3531320}}


\bibitem[Santana et~al\mbox{.}(2023)]%
        {Santana2023COVID}
\bibfield{author}{\bibinfo{person}{Clodomir Santana}, \bibinfo{person}{Federico
  Botta}, \bibinfo{person}{Hugo Barbosa}, \bibinfo{person}{Filippo Privitera},
  \bibinfo{person}{Ronaldo Menezes}, {and} \bibinfo{person}{Riccardo
  Di~Clemente}.} \bibinfo{year}{2023}\natexlab{}.
\newblock \showarticletitle{COVID-19 is linked to changes in the time–space
  dimension of human mobility}.
\newblock \bibinfo{journal}{\emph{Nature Human Behaviour}} \bibinfo{volume}{7},
  \bibinfo{number}{10} (\bibinfo{date}{July} \bibinfo{year}{2023}),
  \bibinfo{pages}{1729--1739}.
\newblock
\showISSN{2397-3374}
\href{https://doi.org/10.1038/s41562-023-01660-3}{doi:\nolinkurl{10.1038/s41562-023-01660-3}}


\bibitem[Septiandri et~al\mbox{.}(2023)]%
        {Septiandri2023WEIRD}
\bibfield{author}{\bibinfo{person}{Ali~Akbar Septiandri},
  \bibinfo{person}{Marios Constantinides}, \bibinfo{person}{Mohammad Tahaei},
  {and} \bibinfo{person}{Daniele Quercia}.} \bibinfo{year}{2023}\natexlab{}.
\newblock \showarticletitle{WEIRD FAccTs: How Western, Educated,
  Industrialized, Rich, and Democratic is FAccT?}. In
  \bibinfo{booktitle}{\emph{ACM Conference on Fairness, Accountability, and
  Transparency}} \emph{(\bibinfo{series}{FAccT'23})}. \bibinfo{publisher}{ACM},
  \bibinfo{address}{New York, NY, USA}, \bibinfo{pages}{160--171}.
\newblock
\showISBNx{9798400701924}
\href{https://doi.org/10.1145/3593013.3593985}{doi:\nolinkurl{10.1145/3593013.3593985}}


\bibitem[Shang et~al\mbox{.}(2023)]%
        {10.1007/978-3-031-23844-4_26}
\bibfield{author}{\bibinfo{person}{Yewei Shang}, \bibinfo{person}{Montserrat
  Pallares-Barbera}, {and} \bibinfo{person}{Francesc Romagosa}.}
  \bibinfo{year}{2023}\natexlab{}.
\newblock \showarticletitle{Coordination Mechanism of Stakeholders in Tourism
  Destinations: A Social Network Analysis Exploration}. In
  \bibinfo{booktitle}{\emph{New Perspectives and Paradigms in Applied Economics
  and Business}}. \bibinfo{publisher}{Springer}, \bibinfo{address}{Cham},
  \bibinfo{pages}{363--381}.
\newblock
\showISBNx{9783031238444}
\showISSN{2198-7254}
\href{https://doi.org/10.1007/978-3-031-23844-4_26}{doi:\nolinkurl{10.1007/978-3-031-23844-4_26}}


\bibitem[Smeral(2019)]%
        {smeral2019overcrowding}
\bibfield{author}{\bibinfo{person}{Egon Smeral}.}
  \bibinfo{year}{2019}\natexlab{}.
\newblock \showarticletitle{Overcrowding of tourism destinations: Some
  suggestions for a solution}.
\newblock In \bibinfo{booktitle}{\emph{Overtourism}}.
  \bibinfo{publisher}{Routledge}, \bibinfo{pages}{163--173}.
\newblock


\bibitem[Song et~al\mbox{.}(2025)]%
        {Song2025Accurate}
\bibfield{author}{\bibinfo{person}{Xiaoyu Song}, \bibinfo{person}{Zhizhong
  Liu}, \bibinfo{person}{Lingqiang Meng}, \bibinfo{person}{Dianhui Chu},
  \bibinfo{person}{Jian Yu}, {and} \bibinfo{person}{Quan~Z. Sheng}.}
  \bibinfo{year}{2025}\natexlab{}.
\newblock \showarticletitle{Accurate POI recommendation for random groups with
  improved graph neural networks and a multi-negotiation model}.
\newblock \bibinfo{journal}{\emph{Scientific Reports}} \bibinfo{volume}{15},
  \bibinfo{number}{1} (\bibinfo{date}{March} \bibinfo{year}{2025}).
\newblock
\showISSN{2045-2322}
\href{https://doi.org/10.1038/s41598-025-91805-3}{doi:\nolinkurl{10.1038/s41598-025-91805-3}}


\bibitem[Soumm et~al\mbox{.}(2023)]%
        {DBLP:conf/wacv/SoummPD23}
\bibfield{author}{\bibinfo{person}{Micha{\"{e}}l Soumm},
  \bibinfo{person}{Adrian Popescu}, {and} \bibinfo{person}{Bertrand
  Delezoide}.} \bibinfo{year}{2023}\natexlab{}.
\newblock \showarticletitle{Vis2Rec: {A} Large-Scale Visual Dataset for Visit
  Recommendation}. In \bibinfo{booktitle}{\emph{{IEEE/CVF} Winter Conference on
  Applications of Computer Vision}}. \bibinfo{publisher}{{IEEE}},
  \bibinfo{pages}{2986--2996}.
\newblock
\href{https://doi.org/10.1109/WACV56688.2023.00300}{doi:\nolinkurl{10.1109/WACV56688.2023.00300}}


\bibitem[Staab et~al\mbox{.}(2002)]%
        {DBLP:journals/expert/StaabWRZGFPK02}
\bibfield{author}{\bibinfo{person}{Steffen Staab}, \bibinfo{person}{Hannes
  Werthner}, \bibinfo{person}{Francesco Ricci}, \bibinfo{person}{Alexander
  Zipf}, \bibinfo{person}{Ulrike Gretzel}, \bibinfo{person}{Daniel~R.
  Fesenmaier}, \bibinfo{person}{C{\'{e}}cile Paris}, {and}
  \bibinfo{person}{Craig~A. Knoblock}.} \bibinfo{year}{2002}\natexlab{}.
\newblock \showarticletitle{Intelligent Systems for Tourism}.
\newblock \bibinfo{journal}{\emph{IEEE Intelligent Systems}}
  \bibinfo{volume}{17}, \bibinfo{number}{6} (\bibinfo{year}{2002}),
  \bibinfo{pages}{53--64}.
\newblock
\href{https://doi.org/10.1109/MIS.2002.1134362}{doi:\nolinkurl{10.1109/MIS.2002.1134362}}


\bibitem[Stankov and Gretzel(2020)]%
        {DBLP:journals/jitt/StankovG20}
\bibfield{author}{\bibinfo{person}{Ugljesa Stankov} {and}
  \bibinfo{person}{Ulrike Gretzel}.} \bibinfo{year}{2020}\natexlab{}.
\newblock \showarticletitle{Tourism 4.0 technologies and tourist experiences: a
  human-centered design perspective}.
\newblock \bibinfo{journal}{\emph{Information Technology \& Tourism}}
  \bibinfo{volume}{22}, \bibinfo{number}{3} (\bibinfo{year}{2020}),
  \bibinfo{pages}{477--488}.
\newblock
\href{https://doi.org/10.1007/S40558-020-00186-Y}{doi:\nolinkurl{10.1007/S40558-020-00186-Y}}


\bibitem[Sun et~al\mbox{.}(2021)]%
        {DBLP:conf/ijcai/Sun00ZC021}
\bibfield{author}{\bibinfo{person}{Huimin Sun}, \bibinfo{person}{Jiajie Xu},
  \bibinfo{person}{Kai Zheng}, \bibinfo{person}{Pengpeng Zhao},
  \bibinfo{person}{Pingfu Chao}, {and} \bibinfo{person}{Xiaofang Zhou}.}
  \bibinfo{year}{2021}\natexlab{}.
\newblock \showarticletitle{{MFNP:} {A} Meta-optimized Model for Few-shot Next
  {POI} Recommendation}. In \bibinfo{booktitle}{\emph{30th International Joint
  Conference on Artificial ntelligence}} \emph{(\bibinfo{series}{IJCAI'21})}.
  \bibinfo{publisher}{International Joint Conferences on Artificial
  Intelligence Organization}, \bibinfo{pages}{3017--3023}.
\newblock
\href{https://doi.org/10.24963/IJCAI.2021/415}{doi:\nolinkurl{10.24963/IJCAI.2021/415}}


\bibitem[Thomee et~al\mbox{.}(2016)]%
        {DBLP:journals/cacm/ThomeeSFENPBL16}
\bibfield{author}{\bibinfo{person}{Bart Thomee}, \bibinfo{person}{David~A.
  Shamma}, \bibinfo{person}{Gerald Friedland}, \bibinfo{person}{Benjamin
  Elizalde}, \bibinfo{person}{Karl Ni}, \bibinfo{person}{Douglas Poland},
  \bibinfo{person}{Damian Borth}, {and} \bibinfo{person}{Li{-}Jia Li}.}
  \bibinfo{year}{2016}\natexlab{}.
\newblock \showarticletitle{{YFCC100M:} the new data in multimedia research}.
\newblock \bibinfo{journal}{\emph{Commun. ACM}} \bibinfo{volume}{59},
  \bibinfo{number}{2} (\bibinfo{year}{2016}), \bibinfo{pages}{64--73}.
\newblock
\href{https://doi.org/10.1145/2812802}{doi:\nolinkurl{10.1145/2812802}}


\bibitem[Tintarev and Masthoff(2022)]%
        {Tintarev2022Beyond}
\bibfield{author}{\bibinfo{person}{Nava Tintarev} {and} \bibinfo{person}{Judith
  Masthoff}.} \bibinfo{year}{2022}\natexlab{}.
\newblock \showarticletitle{Beyond Explaining Single Item Recommendations}.
\newblock In \bibinfo{booktitle}{\emph{Recommender Systems Handbook}},
  \bibfield{editor}{\bibinfo{person}{Francesco Ricci}, \bibinfo{person}{Lior
  Rokach}, {and} \bibinfo{person}{Bracha Shapira}} (Eds.).
  \bibinfo{publisher}{Springer}, \bibinfo{address}{New York, NY},
  \bibinfo{pages}{711--756}.
\newblock
\showISBNx{978-1-0716-2197-4}
\href{https://doi.org/10.1007/978-1-0716-2197-4_19}{doi:\nolinkurl{10.1007/978-1-0716-2197-4_19}}


\bibitem[Tonetto et~al\mbox{.}(2024)]%
        {tonetto2024ethicalprivacyconsiderationslocation}
\bibfield{author}{\bibinfo{person}{Leonardo Tonetto}, \bibinfo{person}{Pauline
  Kister}, \bibinfo{person}{Nitinder Mohan}, {and} \bibinfo{person}{Jörg
  Ott}.} \bibinfo{year}{2024}\natexlab{}.
\newblock \bibinfo{title}{Ethical and Privacy Considerations with Location
  Based Data Research}.
\newblock
\href{https://doi.org/10.48550/arxiv.2403.05558}{doi:\nolinkurl{10.48550/arxiv.2403.05558}}
\showeprint[arxiv]{2403.05558}~[cs.CY]


\bibitem[Trattner et~al\mbox{.}(2018)]%
        {Trattner2018Investigating}
\bibfield{author}{\bibinfo{person}{Christoph Trattner},
  \bibinfo{person}{Alexander Oberegger}, \bibinfo{person}{Leandro~Balby
  Marinho}, {and} \bibinfo{person}{Denis Parra}.}
  \bibinfo{year}{2018}\natexlab{}.
\newblock \showarticletitle{Investigating the Utility of the Weather Context
  for Point of Interest Recommendations}.
\newblock \bibinfo{journal}{\emph{Information Technology \& Tourism}}
  \bibinfo{volume}{19}, \bibinfo{number}{1-4} (\bibinfo{year}{2018}),
  \bibinfo{pages}{117--150}.
\newblock
\href{https://doi.org/10.1007/s40558-017-0100-9}{doi:\nolinkurl{10.1007/s40558-017-0100-9}}


\bibitem[Vathis et~al\mbox{.}(2023)]%
        {DBLP:journals/cor/VathisKPG23}
\bibfield{author}{\bibinfo{person}{Nikolaos Vathis},
  \bibinfo{person}{Charalampos Konstantopoulos}, \bibinfo{person}{Grammati~E.
  Pantziou}, {and} \bibinfo{person}{Damianos Gavalas}.}
  \bibinfo{year}{2023}\natexlab{}.
\newblock \showarticletitle{The Vacation Planning Problem: {A} multi-level
  clustering-based metaheuristic approach}.
\newblock \bibinfo{journal}{\emph{Computers \& Operations Research and their
  Application to Problems of World Concern}}  \bibinfo{volume}{150}
  (\bibinfo{year}{2023}), \bibinfo{pages}{106083}.
\newblock
\href{https://doi.org/10.1016/J.COR.2022.106083}{doi:\nolinkurl{10.1016/J.COR.2022.106083}}


\bibitem[Vecchia et~al\mbox{.}(2023)]%
        {VecchiaMQB23}
\bibfield{author}{\bibinfo{person}{Anna~Dalla Vecchia}, \bibinfo{person}{Sara
  Migliorini}, \bibinfo{person}{Elisa Quintarelli}, {and}
  \bibinfo{person}{Alberto Belussi}.} \bibinfo{year}{2023}\natexlab{}.
\newblock \showarticletitle{A Context-Aware Recommendation System with a
  Crowding Forecaster}. In \bibinfo{booktitle}{\emph{31st Symposium of Advanced
  Database Systems}} \emph{(\bibinfo{series}{{CEUR} Workshop Proceedings},
  Vol.~\bibinfo{volume}{3478})}. \bibinfo{publisher}{CEUR-WS.org},
  \bibinfo{pages}{632--640}.
\newblock


\bibitem[Vecchia et~al\mbox{.}(2024)]%
        {VecchiaMQGB24}
\bibfield{author}{\bibinfo{person}{Anna~Dalla Vecchia}, \bibinfo{person}{Sara
  Migliorini}, \bibinfo{person}{Elisa Quintarelli}, \bibinfo{person}{Mauro
  Gambini}, {and} \bibinfo{person}{Alberto Belussi}.}
  \bibinfo{year}{2024}\natexlab{}.
\newblock \showarticletitle{Promoting sustainable tourism by recommending
  sequences of attractions with deep reinforcement learning}.
\newblock \bibinfo{journal}{\emph{Information Technology \& Tourism}}
  \bibinfo{volume}{26}, \bibinfo{number}{3} (\bibinfo{year}{2024}),
  \bibinfo{pages}{449--484}.
\newblock
\href{https://doi.org/10.1007/S40558-024-00288-X}{doi:\nolinkurl{10.1007/S40558-024-00288-X}}


\bibitem[Wang et~al\mbox{.}(2024a)]%
        {Wang21102024}
\bibfield{author}{\bibinfo{person}{Bin Wang}, \bibinfo{person}{Tiantian Dong},
  \bibinfo{person}{Yunyao Liu}, \bibinfo{person}{Jay Kandampully}, {and}
  \bibinfo{person}{Zhanlu~Tang and}.} \bibinfo{year}{2024}\natexlab{a}.
\newblock \showarticletitle{Males or females in solo or group travel: how do
  they impact travel intentions of potential tourists with different
  self-construals?}
\newblock \bibinfo{journal}{\emph{Current Issues in Tourism}}
  (\bibinfo{year}{2024}), \bibinfo{pages}{1--20}.
\newblock
\href{https://doi.org/10.1080/13683500.2024.2417713}{doi:\nolinkurl{10.1080/13683500.2024.2417713}}


\bibitem[Wang et~al\mbox{.}(2016)]%
        {DBLP:conf/icwsm/WangSZZ16}
\bibfield{author}{\bibinfo{person}{Gang Wang}, \bibinfo{person}{Sarita~Yardi
  Schoenebeck}, \bibinfo{person}{Haitao Zheng}, {and} \bibinfo{person}{Ben~Y.
  Zhao}.} \bibinfo{year}{2016}\natexlab{}.
\newblock \showarticletitle{"Will Check-in for Badges": Understanding Bias and
  Misbehavior on Location-Based Social Networks}. In
  \bibinfo{booktitle}{\emph{Tenth International Conference on Web and Social
  Media}}. \bibinfo{publisher}{{AAAI}}, \bibinfo{pages}{417--426}.
\newblock


\bibitem[Wang et~al\mbox{.}(2024b)]%
        {DBLP:conf/recsys/WangX0GCX24}
\bibfield{author}{\bibinfo{person}{Shirui Wang}, \bibinfo{person}{Bohan Xie},
  \bibinfo{person}{Ling Ding}, \bibinfo{person}{Xiaoying Gao},
  \bibinfo{person}{Jianting Chen}, {and} \bibinfo{person}{Yang Xiang}.}
  \bibinfo{year}{2024}\natexlab{b}.
\newblock \showarticletitle{SeCor: Aligning Semantic and Collaborative
  Representations by Large Language Models for Next-Point-of-Interest
  Recommendations}. In \bibinfo{booktitle}{\emph{18th {ACM} Conference on
  Recommender Systems}}. \bibinfo{publisher}{{ACM}}, \bibinfo{pages}{1--11}.
\newblock
\href{https://doi.org/10.1145/3640457.3688124}{doi:\nolinkurl{10.1145/3640457.3688124}}


\bibitem[Wang et~al\mbox{.}(2024c)]%
        {WangZWR24}
\bibfield{author}{\bibinfo{person}{Shoujin Wang}, \bibinfo{person}{Xiuzhen
  Zhang}, \bibinfo{person}{Yan Wang}, {and} \bibinfo{person}{Francesco Ricci}.}
  \bibinfo{year}{2024}\natexlab{c}.
\newblock \showarticletitle{Trustworthy Recommender Systems}.
\newblock \bibinfo{journal}{\emph{ACM Transactions on Intelligent Systems and
  Technology}} \bibinfo{volume}{15}, \bibinfo{number}{4}
  (\bibinfo{year}{2024}), \bibinfo{pages}{84:1--84:20}.
\newblock
\href{https://doi.org/10.1145/3627826}{doi:\nolinkurl{10.1145/3627826}}


\bibitem[Wang et~al\mbox{.}(2019)]%
        {DBLP:conf/icml/WangZ0Q19}
\bibfield{author}{\bibinfo{person}{Xiaojie Wang}, \bibinfo{person}{Rui Zhang},
  \bibinfo{person}{Yu Sun}, {and} \bibinfo{person}{Jianzhong Qi}.}
  \bibinfo{year}{2019}\natexlab{}.
\newblock \showarticletitle{Doubly Robust Joint Learning for Recommendation on
  Data Missing Not at Random}. In \bibinfo{booktitle}{\emph{36th International
  Conference on Machine Learning}} \emph{(\bibinfo{series}{Proceedings of
  Machine Learning Research}, Vol.~\bibinfo{volume}{97})}.
  \bibinfo{publisher}{{PMLR}}, \bibinfo{pages}{6638--6647}.
\newblock


\bibitem[Wang et~al\mbox{.}(2025)]%
        {DBLP:journals/jitt/WangHJ25}
\bibfield{author}{\bibinfo{person}{Zehui Wang}, \bibinfo{person}{Wolfram
  H{\"{o}}pken}, {and} \bibinfo{person}{Dietmar Jannach}.}
  \bibinfo{year}{2025}\natexlab{}.
\newblock \showarticletitle{A survey on point-of-interest recommendations
  leveraging heterogeneous data}.
\newblock \bibinfo{journal}{\emph{Information Technology \& Tourism}}
  \bibinfo{volume}{27}, \bibinfo{number}{1} (\bibinfo{year}{2025}),
  \bibinfo{pages}{29--73}.
\newblock
\href{https://doi.org/10.1007/S40558-024-00301-3}{doi:\nolinkurl{10.1007/S40558-024-00301-3}}


\bibitem[Werneck et~al\mbox{.}(2021)]%
        {DBLP:journals/eswa/WerneckSSPMR21}
\bibfield{author}{\bibinfo{person}{Heitor Werneck}, \bibinfo{person}{Rodrigo
  Santos}, \bibinfo{person}{N{\'{\i}}collas Silva}, \bibinfo{person}{Adriano
  C.~M. Pereira}, \bibinfo{person}{Fernando Mour{\~{a}}o}, {and}
  \bibinfo{person}{Leonardo Rocha}.} \bibinfo{year}{2021}\natexlab{}.
\newblock \showarticletitle{Effective and diverse {POI} recommendations through
  complementary diversification models}.
\newblock \bibinfo{journal}{\emph{Expert Systems with Applications}}
  \bibinfo{volume}{175} (\bibinfo{year}{2021}), \bibinfo{pages}{114775}.
\newblock
\href{https://doi.org/10.1016/J.ESWA.2021.114775}{doi:\nolinkurl{10.1016/J.ESWA.2021.114775}}


\bibitem[Werthner and Klein(1999)]%
        {Werthner01011999}
\bibfield{author}{\bibinfo{person}{Hannes Werthner} {and}
  \bibinfo{person}{Stefan Klein}.} \bibinfo{year}{1999}\natexlab{}.
\newblock \showarticletitle{ICT and the Changing Landscape of Global Tourism
  Distribution}.
\newblock \bibinfo{journal}{\emph{Electronic Markets}} \bibinfo{volume}{9},
  \bibinfo{number}{4} (\bibinfo{year}{1999}), \bibinfo{pages}{256--262}.
\newblock
\href{https://doi.org/10.1080/101967899358941}{doi:\nolinkurl{10.1080/101967899358941}}


\bibitem[Xiao et~al\mbox{.}(2024)]%
        {DBLP:conf/ijcai/Xiao00XWY24}
\bibfield{author}{\bibinfo{person}{Yanan Xiao}, \bibinfo{person}{Lu Jiang},
  \bibinfo{person}{Kunpeng Liu}, \bibinfo{person}{Yuanbo Xu},
  \bibinfo{person}{Pengyang Wang}, {and} \bibinfo{person}{Minghao Yin}.}
  \bibinfo{year}{2024}\natexlab{}.
\newblock \showarticletitle{Hierarchical Reinforcement Learning for Point of
  Interest Recommendation}. In \bibinfo{booktitle}{\emph{33rd International
  Joint Conference on Artificial Intelligence}}
  \emph{(\bibinfo{series}{IJCAI'24})}. \bibinfo{publisher}{International Joint
  Conferences on Artificial Intelligence Organization},
  \bibinfo{pages}{2460--2468}.
\newblock
\href{https://doi.org/10.24963/ijcai.2024/272}{doi:\nolinkurl{10.24963/ijcai.2024/272}}


\bibitem[Xu et~al\mbox{.}(2023)]%
        {Xu2023Urban}
\bibfield{author}{\bibinfo{person}{Yanyan Xu}, \bibinfo{person}{Luis~E. Olmos},
  \bibinfo{person}{David Mateo}, \bibinfo{person}{Alberto Hernando},
  \bibinfo{person}{Xiaokang Yang}, {and} \bibinfo{person}{Marta~C. González}.}
  \bibinfo{year}{2023}\natexlab{}.
\newblock \showarticletitle{Urban dynamics through the lens of human mobility}.
\newblock \bibinfo{journal}{\emph{Nature Computational Science}}
  \bibinfo{volume}{3}, \bibinfo{number}{7} (\bibinfo{date}{July}
  \bibinfo{year}{2023}), \bibinfo{pages}{611--620}.
\newblock
\showISSN{2662-8457}
\href{https://doi.org/10.1038/s43588-023-00484-5}{doi:\nolinkurl{10.1038/s43588-023-00484-5}}


\bibitem[Yabe et~al\mbox{.}(2024)]%
        {DBLP:data/10/YabeTSSSMP24}
\bibfield{author}{\bibinfo{person}{Takahiro Yabe}, \bibinfo{person}{Kota
  Tsubouchi}, \bibinfo{person}{Toru Shimizu}, \bibinfo{person}{Yoshihide
  Sekimoto}, \bibinfo{person}{Kaoru Sezaki}, \bibinfo{person}{Esteban Moro},
  {and} \bibinfo{person}{Alex Pentland}.} \bibinfo{year}{2024}\natexlab{}.
\newblock \bibinfo{title}{YJMob100K: City-Scale and Longitudinal Dataset of
  Anonymized Human Mobility Trajectories}.
\newblock
\href{https://doi.org/10.5281/ZENODO.10836269}{doi:\nolinkurl{10.5281/ZENODO.10836269}}


\bibitem[Yalcin and Bilge(2022)]%
        {DBLP:journals/ipm/YalcinB22}
\bibfield{author}{\bibinfo{person}{Emre Yalcin} {and} \bibinfo{person}{Alper
  Bilge}.} \bibinfo{year}{2022}\natexlab{}.
\newblock \showarticletitle{Evaluating unfairness of popularity bias in
  recommender systems: {A} comprehensive user-centric analysis}.
\newblock \bibinfo{journal}{\emph{Information Processing and Management}}
  \bibinfo{volume}{59}, \bibinfo{number}{6} (\bibinfo{year}{2022}),
  \bibinfo{pages}{103100}.
\newblock
\href{https://doi.org/10.1016/J.IPM.2022.103100}{doi:\nolinkurl{10.1016/J.IPM.2022.103100}}


\bibitem[Yan et~al\mbox{.}(2023)]%
        {Yan2023Personalized}
\bibfield{author}{\bibinfo{person}{An Yan}, \bibinfo{person}{Zhankui He},
  \bibinfo{person}{Jiacheng Li}, \bibinfo{person}{Tianyang Zhang}, {and}
  \bibinfo{person}{Julian McAuley}.} \bibinfo{year}{2023}\natexlab{}.
\newblock \showarticletitle{Personalized Showcases: Generating Multi-Modal
  Explanations for Recommendations}. In \bibinfo{booktitle}{\emph{46th
  International ACM SIGIR Conference on Research and Development in Information
  Retrieval}} \emph{(\bibinfo{series}{SIGIR'23})}. \bibinfo{publisher}{ACM},
  \bibinfo{address}{New York, NY, USA}, \bibinfo{pages}{2251--2255}.
\newblock
\showISBNx{9781450394086}
\href{https://doi.org/10.1145/3539618.3592036}{doi:\nolinkurl{10.1145/3539618.3592036}}


\bibitem[Zhang and Wang(2016)]%
        {DBLP:journals/kais/ZhangW16}
\bibfield{author}{\bibinfo{person}{Chenyi Zhang} {and} \bibinfo{person}{Ke
  Wang}.} \bibinfo{year}{2016}\natexlab{}.
\newblock \showarticletitle{{POI} recommendation through cross-region
  collaborative filtering}.
\newblock \bibinfo{journal}{\emph{Knowledge and Information Systems}}
  \bibinfo{volume}{46}, \bibinfo{number}{2} (\bibinfo{year}{2016}),
  \bibinfo{pages}{369--387}.
\newblock
\href{https://doi.org/10.1007/S10115-015-0825-8}{doi:\nolinkurl{10.1007/S10115-015-0825-8}}


\bibitem[Zhang et~al\mbox{.}(2023)]%
        {DBLP:journals/tnn/ZhangSZWX23}
\bibfield{author}{\bibinfo{person}{Lu Zhang}, \bibinfo{person}{Zhu Sun},
  \bibinfo{person}{Jie Zhang}, \bibinfo{person}{Yiwen Wu}, {and}
  \bibinfo{person}{Yunwen Xia}.} \bibinfo{year}{2023}\natexlab{}.
\newblock \showarticletitle{Conversation-Based Adaptive Relational Translation
  Method for Next {POI} Recommendation With Uncertain Check-Ins}.
\newblock \bibinfo{journal}{\emph{{IEEE} Transactions on Neural Networks and
  Learning Systems}} \bibinfo{volume}{34}, \bibinfo{number}{10}
  (\bibinfo{year}{2023}), \bibinfo{pages}{7810--7823}.
\newblock
\href{https://doi.org/10.1109/TNNLS.2022.3146443}{doi:\nolinkurl{10.1109/TNNLS.2022.3146443}}


\bibitem[Zheng et~al\mbox{.}(2010)]%
        {DBLP:conf/www/ZhengZXY10}
\bibfield{author}{\bibinfo{person}{Vincent~Wenchen Zheng}, \bibinfo{person}{Yu
  Zheng}, \bibinfo{person}{Xing Xie}, {and} \bibinfo{person}{Qiang Yang}.}
  \bibinfo{year}{2010}\natexlab{}.
\newblock \showarticletitle{Collaborative location and activity recommendations
  with {GPS} history data}. In \bibinfo{booktitle}{\emph{19th International
  Conference on World Wide Web}}. \bibinfo{publisher}{{ACM}},
  \bibinfo{pages}{1029--1038}.
\newblock
\href{https://doi.org/10.1145/1772690.1772795}{doi:\nolinkurl{10.1145/1772690.1772795}}


\end{thebibliography}
\end{document}